\documentclass[11pt]{article}

\usepackage[utf8]{inputenc}
\usepackage{amssymb,verbatim,epsfig}
\usepackage{amsmath,setspace}
\usepackage{siunitx}
\usepackage{multicol}
\usepackage{multirow}
\usepackage{lipsum}  
\usepackage[style = numeric-comp, sorting = none, backend=biber,maxbibnames=99]{biblatex}
\DeclareSourcemap{
  \maps[datatype=bibtex]{
    \map{
      \step[fieldset=issn, null]
    }
  }
}
\usepackage{vcell}
\usepackage{url}
\usepackage{wrapfig}
\usepackage[capitalize]{cleveref}
\usepackage{amssymb, amsmath, verbatim, epsfig}
\usepackage{graphicx}
\usepackage{setspace}
\usepackage{longtable}
\usepackage{lineno}
\usepackage[margin=1in]{geometry} 
\usepackage[small,bf]{caption}
\usepackage{textcomp}
\usepackage{gensymb}

\newcommand{\volume}{{\ooalign{\hfil$V$\hfil\cr\kern0.08em--\hfil\cr}}}
\spacing{1} 
\bibliography{main}
\usepackage{appendix}

\title{\bf 
\vspace{-0.75in}
Cavitation induced by pulsed and continuous-wave \\ fiber lasers in confinement} 

\author{ Jelle J. Schoppink$^{1*}$, Jan Krizek$^2$, Christophe Moser$^2$ \& 
David Fernandez Rivas$^1$, \\ 
\small$^1$Mesoscale Chemical Systems group, MESA+ Institute and Faculty of Science and Technology, \\  
\small University  of Twente, P.O. Box 217, 7500 AE Enschede, the Netherlands. $^2$Laboratory of \\
\small  Applied Photonics Devices, École Polytechnique Fédérale de Lausanne (EPFL), Station 17, \\
\small 1015 Lausanne, Switzerland. $^*$ Corresponding author: j.j.schoppink@utwente.nl}

\date{\today}
\begin{document} \maketitle 
\begin{abstract}
 Bubbles generated with lasers under confinement have been investigated for their potential use as the driving mechanism for liquid micro-jets in various microfluidic devices, such as needle-free jet injectors. 
 Here, we report on the study of bubble formation by a continuous-wave (CW) and a pulsed laser inside an open-ended microfluidic capillary. 
 This results in a direct comparison between bubbles generated by laser sources emitting light in different time scales (ms and ns). 
 The bubble kinetics represents an important parameter because it determines the available kinetic energy for a subsequent liquid jet.

 We show that the bubble growth rate increases linearly with the delivered energy for both the CW and the pulsed laser. Experiments show that at equal absorption coefficient, the bubble growth for both lasers is similar, which indicates that they can be used interchangeably for a jet generation. However, bubbles generated by a CW laser require more optical energy, which is due to heat dissipation. Furthermore, the bubbles generated by the CW laser show a slightly larger variation in size and growth rate for identical initial conditions, which we attribute to the stochastic nature of thermocavitation. 
\end{abstract} 

\maketitle
\vspace{0.1 in}
\noindent{\bf Keywords:} [Vapor bubble, Cavitation, Pulsed laser, CW laser, Jet injection, Thermocavitation, Microfluidic confinement]
\section{Introduction}\label{sec: intro}
Laser cavitation is widely used for medical applications~\cite{Peng2008}, including ablation of biological tissue~\cite{Vogel2003} and lithotripsy~\cite{Fried2018a,Marks2007}. Lately, lasers have also been investigated for their use in laser-actuated needle-free jet injectors (NFJIs)~\cite{Schoppink2022}. These NFJIs present many advantages over conventional hypodermic needles, such as improved patient compliance~\cite{Daly2020}, reduction in needle waste~\cite{Mitragotri2005}, improved safety for healthcare workers~\cite{Mitragotri2006}, injection of high-viscous liquids~\cite{Williams2016} and better control over injection parameters~\cite{Schoppink2022,McKeage2018a}. 

The working principle of these laser-actuated NFJIs is similar to that of laser-induced forward transfer, where a laser is heating a liquid resulting in the formation of a vapor bubble~\cite{Sopena2017,Santos2020}. For NFJIs, the liquid is contained in an open-ended microfluidic channel, such that the explosively growing bubble pushes the liquid through a small opening, which results in the formation of a fast microfluidic jet with the ability to penetrate the skin. 
Jet velocities are controlled by the delivered energy~\cite{Tagawa2012}, optical beam size~\cite{Krizek2020}, channel size~\cite{Tagawa2012,Berrospe-Rodriguez2016} or the inclusion of a tapered nozzle~\cite{OyarteGalvez2020}. The injection depth into the tissue is proportional to this jet velocity~\cite{Krizek2020}, which can be used to target specific skin layers~\cite{Schoppink2022}.

Although the liquid is heated by the laser, temperature measurements show moderate and temporal temperature increase~\cite{Quinto-Su2014, Banks2019}. Furthermore, degradation studies showed no damage to the drug molecules~\cite{Krizek2020,Krizek2021}. To avoid any risks of degrading the liquid to be injected, some injectors include a membrane to split two chambers~\cite{Han2010,Ham2017}. For these reasons, laser heating of the liquid is not hindering the applicability of jet injection.

Besides the standard concept of laser-actuated jet injection, several additions have been introduced. Instead of using lenses to guide and focus the light, the use of optical fibers allows for smaller and more flexible devices and results in a reproducible beam size~\cite{Krizek2020,Krizek2021}. Furthermore, repetitive jetting results in the desired injection volume and can increase the injection depth~\cite{Jang2014,Ham2017,Krizek2020a,OyarteGalvez2019,Arora2007,Romgens2016}. Finally, the addition of a spacer between the injector fixes the stand-off distance and applies tensile stress on the skin to improve the injection reproducibility. All of these additions show improvements for the jet injector.

\begin{figure}[b!]
    
    \includegraphics[width=0.6\linewidth]{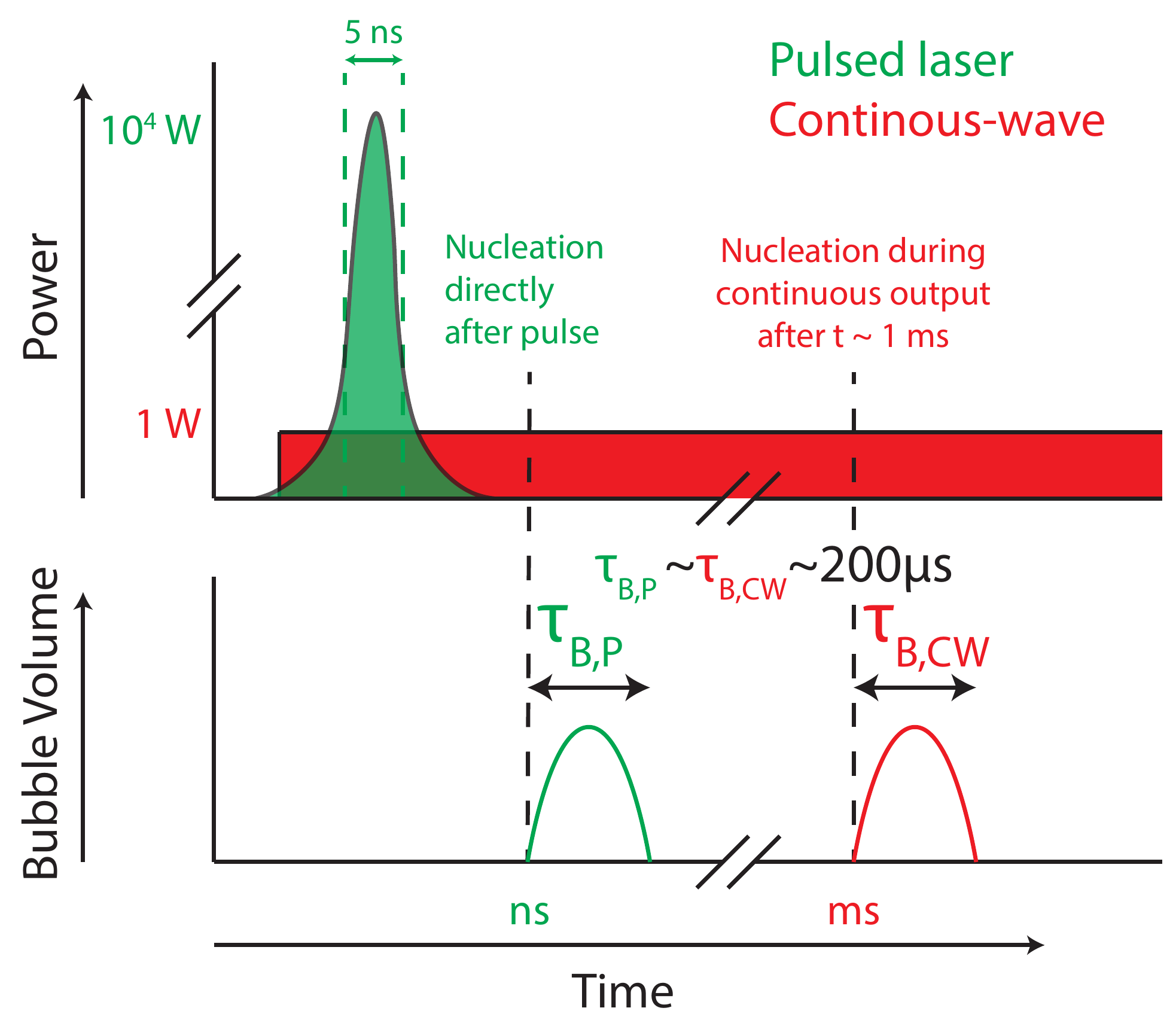}
    \centering
    \captionsetup{width=\linewidth}
    \caption{Schematics indicating the main difference between the timescale of nucleation and power for the pulsed and continuous-wave laser. For the pulsed laser (green), all energy is delivered with a high peak power within 5 ns, after which nucleation occurs and the bubble forms. For the continuous-wave laser (red), the power is switched on and remains working continuously at low power. After some time during this laser illumination ($\tau\sim\textnormal{ms}$, depending on the power), bubble nucleation occurs. Although the timescales are different, the bubbles show similar dynamics and lifetimes ($\tau_\textnormal{P,B}\sim$~$\tau_\textnormal{C,B}\sim$~200~µs)}
    \label{fig: Laser types}
\end{figure}

NFJIs can be actuated by a pulsed laser~\cite{Krizek2020,Ham2017,Robles2020,Rohilla2020,Tagawa2012,Tagawa2013}, or a continuous-wave (CW) laser~\cite{Berrospe-Rodriguez2016,Berrospe-Rodriguez2017, OyarteGalvez2020,Quetzeri-Santiago2021}. Although both laser types can create jets able to penetrate skin-like substrates, a comparison between the two methods is difficult, as all studies use different injector geometries, laser configurations, ejected liquid volumes and disparate substrates for injection~\cite{Schoppink2022}. The difference between the two laser types is in the time of irradiation, as shown in Figure \ref{fig: Laser types}. For the pulsed laser, the bubble forms shortly after the ns pulse with very high peak power ($\approx 0.5-2  GW/cm^2$). The CW laser irradiates continuously and therefore heats the liquid slower (ms). At a certain time after the beginning of radiation ($\tau_C \sim$~ms), there is enough energy for nucleation and subsequently, the bubble forms. The difference in the timescales (ns and ms) between these two lasers has an effect on the absorption of energy, heat and pressure confinement, which will be further discussed in Section \ref{sec: Theory}.

In this manuscript, we investigate laser-actuated cavitation at a fiber tip in an open capillary, and we compare two types of lasers, the ns-pulsed and the CW laser. This comparison provides a better understanding of the energy transfer from the optical energy into the kinetic energy, which is of great importance for the use of laser-actuated cavitation in needle-free jet injectors. 

\section{Theory}\label{sec: Theory}
Figure \ref{fig: Laser types} shows the difference in timescales between nucleation by the pulsed (ns) and CW lasers (ms). The difference in timescale has an effect on three parameters: the absorption of the optical energy, thermal diffusion and potential pressure confinement. These will be explained below.

First of all, the energy absorption by the liquid can be split up into two categories: linear and non-linear. The former depends on the absorption coefficient $\alpha$ of the liquid at the laser wavelength. As linear absorption does not depend on timescale, it is resent in both pulsed and CW laser exposition. In the case of only linear absorption, the optical penetration can be calculated by~\cite{Jacques1993}
\begin{equation}
    \Psi(z) = \Psi_0 \exp(-\alpha z),
\end{equation}
where $\Psi$ is the laser irradiance, $\Psi_0$ is the incident irradiance, and z is the position along the laser beam. Therefore, a typical penetration depth  $d = \frac{1}{\alpha}$ can be defined, at which the irradiance has dropped to $\Psi(d) = \exp(-1) \Psi_0 \approx 0.37 \Psi_0$, which means that approximately 63\% has been absorbed. As linear absorption of water is negligible in the visible and near-infrared ($\lambda <$ 1300~nm)~\cite{Ruru2012}, a dye is typically added to increase the absorption~\cite{Padilla-Martinez2014}. Non-linear absorption may occur at sufficient power densities, which is a result of optical or thermal breakdown \cite{Vogel2003}.
The threshold for this optical breakdown is found to be around $10^{-11}$~W/m$^2$ or $\sim$~500~J/cm$^2$ for laser pulses of 6~ns at a wavelength of 1064~nm~\cite{Vogel1999}. However, the threshold is reduced up to three orders of magnitude when the target has a very high linear absorption coefficient~\cite{Vogel2003}.

Second of all, thermal diffusion plays a significant role when the timescale of nucleation is comparable to the thermal diffusion timescale. The thermal diffusion time $t_d$ is given as~\cite{Paltauf2003}
\begin{equation}\label{eq: thermal diffusion}
    t_d = \frac{\delta^2}{4\kappa},
\end{equation}
where $\kappa$ is the thermal diffusivity of the liquid, and $\delta$ is the typical length scale, which is either the beam diameter or the typical absorption length, whichever is smaller. In the case of heating water ($\kappa~\approx~0.14~\textnormal{mm}^2/\textnormal{s}$) with a beam diameter of 50~µm and absorption length of 100~µm, the thermal diffusion time (over the length of the beam diameter) is approximately 4 ms. Therefore, thermal diffusion does not play a role in the pulsed laser actuation ($\tau_{pulsed} \sim $ ns), but does influence the CW actuation ($\tau_{cw} \sim $ ms). 

However, even for pulsed lasers, heat transfer during the bubble's lifetime should be taken into account. Sun et al. found that the inclusion of heat transfer is required for a numerical model of the growth and collapse of the bubble~\cite{Sun2009}. They also found that the bubble collapse is slower than the growth in microchannels, and such an effect increases with smaller channel sizes. 

Third of all, the rapid heating of the liquid results in thermoelastic stresses in the irradiated volume, as the system tends to reconfigure to a new equilibrium~\cite{Vogel2003}. Pressure confinement may play an important role in defining the threshold for liquid-gas transition \cite{Paltauf1998}. In the case of irradiation from an optical fiber, the finite size of the fiber enhances this effect~\cite{Paltauf1998}. When this heating is sufficiently fast, the pressure is confined within the volume of irradiation near the fiber tip. The time for a pressure wave to travel across the irradiated volume is~\cite{Frenz1996}
\begin{equation}
    t_p = \frac{\delta}{c_\textnormal{water}},
\end{equation}
where $c_\textnormal{water}$ is the speed of sound in water (1480~m/s). For a fiber diameter of 50~µm and absorption length of 100~µm, $t_p~\approx$~34~ns. This means that for pulsed laser actuation, the pressure is mostly confined to the region close to the fiber. In the absence of pressure confinement (when the laser pulse duration is much longer than $t_p$), there is significant thermal expansion of the heated volume during the irradiation. Therefore, the thermoelastic stresses are reduced, and the built-up pressure at bubble formation is reduced~\cite{Vogel2003}.

\section{Experimental methods}\label{sec: methods}

\begin{figure}[b!]
    
    \includegraphics[width=\linewidth]{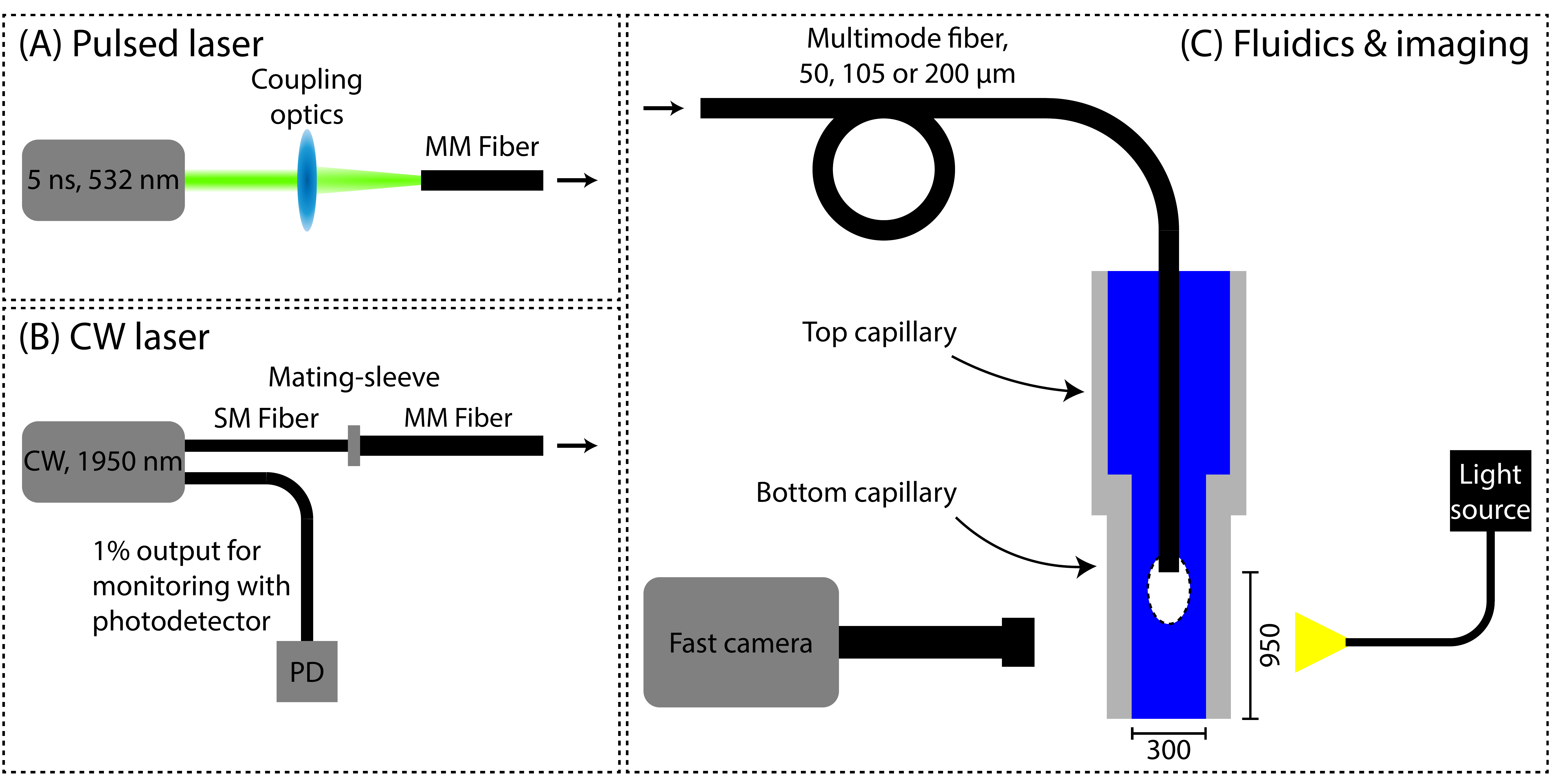}
    \centering
    \captionsetup{width=\linewidth}
    \caption{Experimental setups consisting of a laser, either pulsed (A) or continuous-wave (B), coupled into a multimode optical fiber, which is inserted into a capillary system (C). A high-speed camera (Photron NOVA S6), together with a light source (Schott CV-LS) is used for imaging. The inner diameter of the smaller capillary is 300~µm, and the distance between the fiber tip and the liquid-air interface is 950~µm. Due to the smaller inner diameter of the lower capillary compared to the upper capillary (ID = 1200~µm), surface tension ensures that the lower capillary will constantly be filled after each jet generation.}
    \label{fig: ExperimentalSetup}
\end{figure} 

Figure~\ref{fig: ExperimentalSetup} shows the experimental setups, which consist of a fiber-coupled laser and a microfluidic capillary system. Two laser types are used, a pulsed and a CW laser. For all the experiments, the laser is coupled into a multimode optical fiber with core diameter varying from 50, 105 and 200~µm (Thorlabs FG050LGA, FG105LCA, FG200LEA). This fiber is inserted into a capillary system filled with water. Upon laser illumination, a bubble forms at the laser tip. 

The pulsed laser (Continuum, ML-II) has a pulse duration of 5~ns and a wavelength of 532~nm. The laser light is coupled into the optical fiber using a two-mirror system and a focusing objective. Before and after each experiment, the pulse energy at the output of the optical fiber is measured using an energy sensor (Thorlabs ES111C). The light energy coupled into the fiber varied between 50 and 700~µJ. For the 50 and 105~µm fiber the upper pulse energy limit was approximately 130 and 480~µJ due to the laser-induced damage threshold for the fiber tip. 

The CW laser (BKTel Photonics, HPFL-2-350-FCAPC) has a wavelength of 1950~nm, which was deliberately chosen to match the absorption peak of water ($\alpha~\approx$ 12000~m$^{-1}$~\cite{Ruru2012}). The output power can be varied from 0.2 to 3W. The laser is initially coupled into a single-mode fiber (Corning SMF-28e), which is then connected through a mating sleeve to the respective multimode fiber. The laser also has a secondary fiber output at 1\% of the nominal power, which is connected to a photodetector (Thorlabs DET05D2) to monitor the output power using an oscilloscope (Tektronix MSO2014B).

The multimode fiber is inserted into a capillary system, which consists of two concentric connected capillaries with inner diameters of 1500 (top) and 300~µm (bottom). Initially, the capillary system is completely filled with water, and capillary flow from the larger to the smaller capillary ensures that the 300~µm capillary will be completely filled. The fiber is inserted partially into the 300~µm (bottom) capillary, and has a distance of 950~µm to the end of the capillary (the liquid-air interface).

For the experiments with the 532~nm laser and 105~µm fiber, we increased the absorption with aqueous solutions of Allura Red AC (ARAC, red food dye) with varying concentrations between 5 and 52~mM. The absorption coefficients were measured and compared with values reported in the literature (see Appendix A). For the 50 and 200~µm fiber, only the 10~mM ($\approx~12000~\textnormal{m}^{-1}$) and 52~mM ($\approx~19000~\textnormal{m}^{-1}$) were used. The 10~mM was chosen as the absorption coefficient matches the absorption coefficient of the CW laser experiment in this study, and the 52~mM was chosen as it was used in previous studies~\cite{Krizek2020,Krizek2020a}.

\begin{figure}
\centering

\begin{minipage}{.49\textwidth}
  \centering
  \includegraphics[width=.9\linewidth]{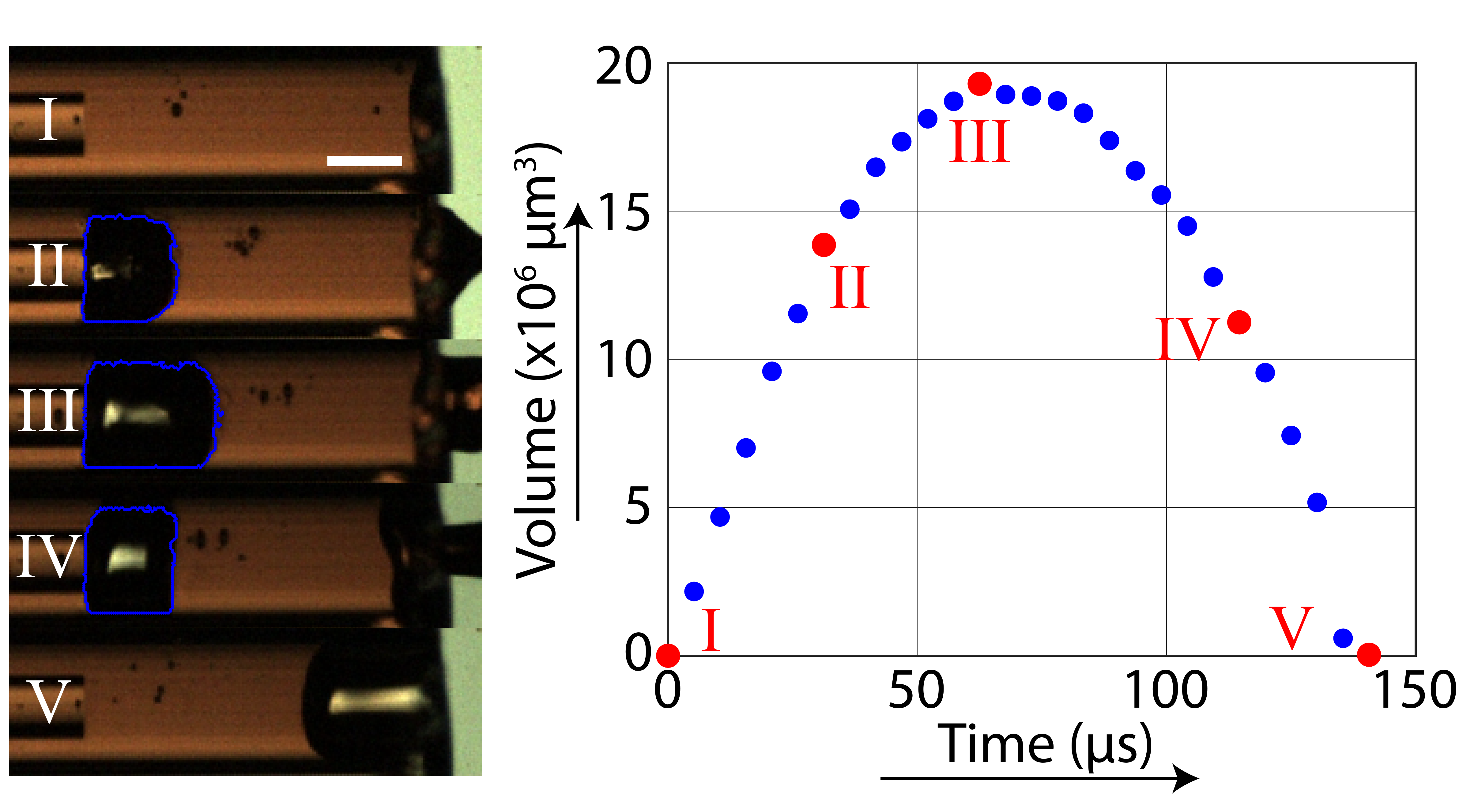}
  \captionsetup{width=.9\linewidth}
  \caption{Bubble volume over time for an experiment with the pulsed laser (d$_{\textrm{fiber}}$ = 105~µm, dye concentration C$_{\textrm{ARAC}}$ = 10~mM, E$_{\textrm{pulse}}$ = 233~µJ). \textbf{Left:} Five images during the experiment, with the tracked bubble in blue. Respectively: just before the laser pulse, during the growth phase, at maximum volume, during the collapse, and after the experiment showing the expelled volume. The white scale bar in the top figure measures 200 µm.  \textbf{Right:} Bubble volume plotted versus time dependence, the red dots corresponds to the 5 images on the left (I-V).}
  \label{fig: BubbleVolumeTime}
\end{minipage}
\vline
\begin{minipage}{.49\textwidth}
  \centering
  \includegraphics[width=.9\linewidth]{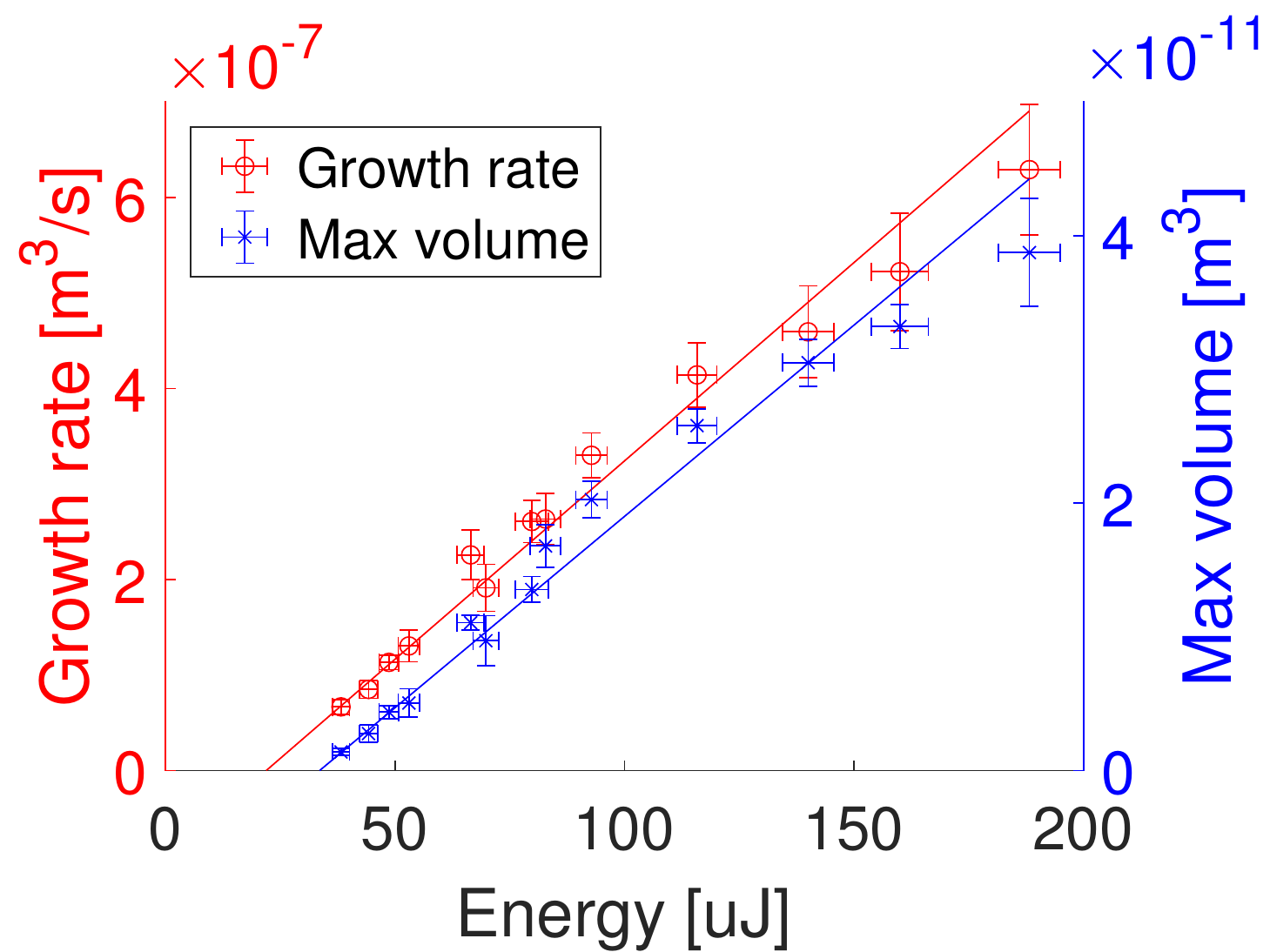}
  \captionsetup{width=.9\linewidth}
  \caption{Maximum bubble volume (blue, right axis) and average volumetric growth rate (red, left axis) for a range of pulse energies with the 105~µm fiber and 52~mM dye concentration. Each data point is an average of at least 6 individual measurements with the error bars indicating standard deviation. Solid lines indicate a linear fit for both variables.}
  \label{fig: Pulsed105 Volume Growth rate}
\end{minipage}
\end{figure}

A Photron NOVA S6 high-speed camera was used in combination with Navitar 12x zoom lens system and a Schott CV-LS light source for visualization of the bubble dynamics. The camera is used at a framerate of 192k fps, a resolution of 256*80 and a pixel size of 5~µm. Figure~\ref{fig: BubbleVolumeTime} shows a few typical images during the bubble lifetime. The images are analyzed with a custom-made MATLAB algorithm, which tracks the bubble over time, and calculates the volume assuming cylindrical symmetry. The average growth rate is calculated as the maximum bubble size divided by the growth time.

\section{Results and discussion}\label{sec: results and discussion}
\subsection{Pulsed laser}\label{sec: pulsed laser}
\subsubsection*{Bubble volume and volumetric growth rate}
Figure~\ref{fig: BubbleVolumeTime}(a) shows typical experimental images of the bubble generated by the pulsed laser. The bubble appears at the fiber tip and grows towards the free surface until it reaches a maximum size and collapses. Occasionally, cavitation bubbles inside the bulk liquid appear due to the negative peak pressure of the shock wave within a few frames after the laser pulse ($\tau~<$~20~µs). As these cavitation bubbles remain small (R~$<10$~µm) and do not affect the main bubble or the jet, we will not discuss them further. These bubbles are further discussed in Ref.~\cite{Hayasaka2017}.

The calculated bubble volume over time is shown in Figure~\ref{fig: BubbleVolumeTime}(b). We can observe parabolic-like behavior, although the growth is typically slightly faster than the collapse ($\approx5-20\%$), which was also observed by Sun et al., both numerically as well as experimentally~\cite{Sun2009}. From the bubble volume curve, the average growth rate is calculated by the maximum size divided by the growth time.

Figure~\ref{fig: Pulsed105 Volume Growth rate} shows the maximum bubble volume and average volumetric growth rate for a range of laser pulse energies. Both the volume and the growth rate show a linear behaviour with the pulse energy, meaning that it is a good indication of the jet velocity, which also increases linearly with pulse energy~\cite{Krizek2020}. In this previous study, it was confirmed that the jet velocity grows linearly with laser energy using the same range of energies, resulting in velocities 0~-~125~m/s~\cite{Krizek2020} and using a dye concentration of (52~mM ARAC).

Interestingly, the offset of the fit, which can be seen as the threshold energy for bubble formation, is not equal for both fits. Therefore, extrapolating this linear relation would indicate that bubbles with zero volume still have a positive growth rate, which wouldn't be possible. We attribute this to the fact that for these very small bubbles, there are errors in bubble detection due to their small size and their short lifetime. These smallest bubbles typically only span 10 by 20 pixels in the image and have a bubble lifetime of approximately 40~ms. Thus, the growth time is only $\sim$~20~ms, which is 4 frames. This means that the actual maximum bubble size may actually be reached between two frames and not captured by the camera. Therefore, we think that the average growth rate is a better parameter to show the bubble dynamics as a function of pulse energy. 

\subsubsection*{Dye concentration effect}
\begin{figure}[t!]
  \centering
  \includegraphics[width=.85\linewidth]{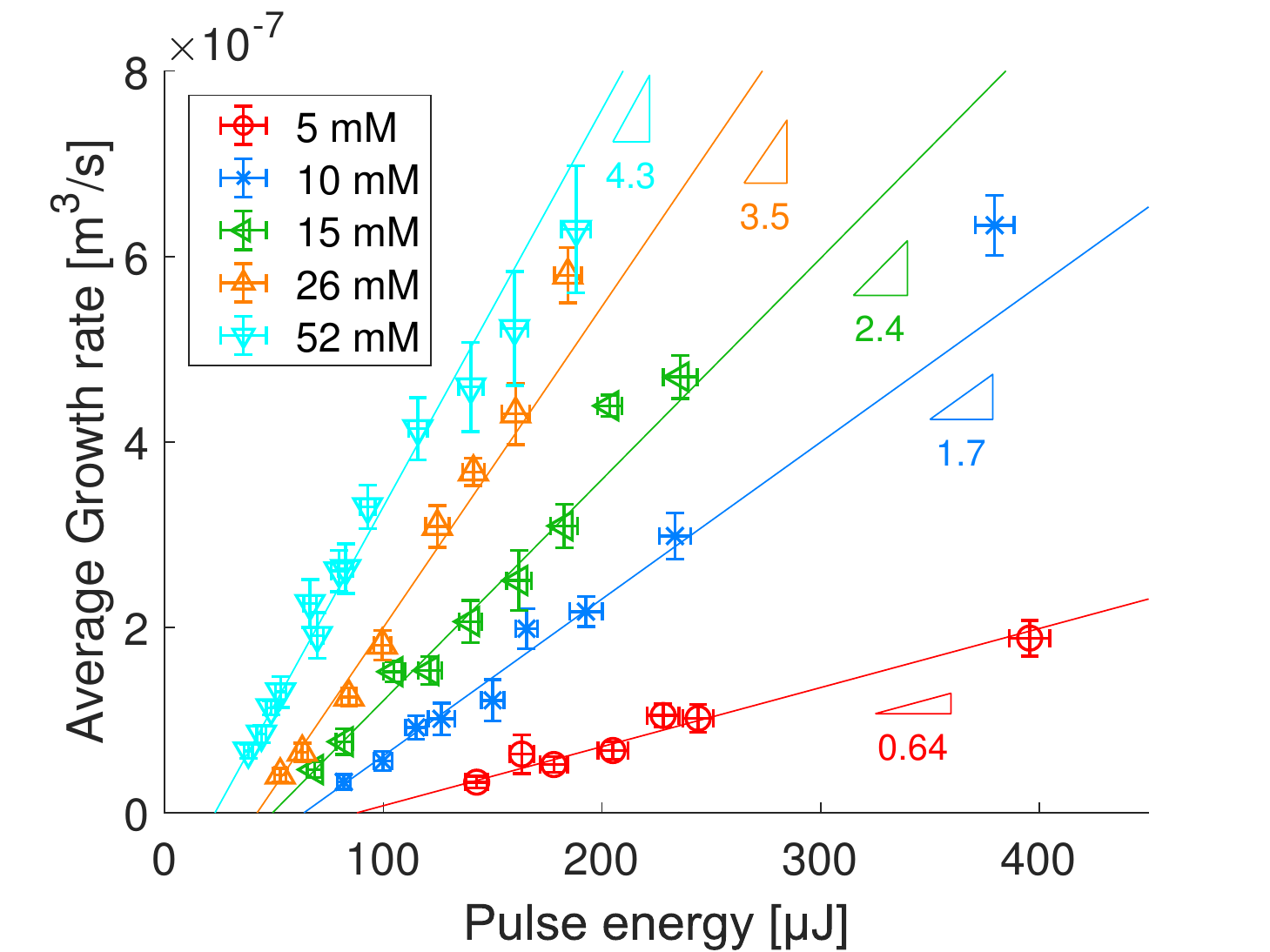}
  \captionsetup{width=.9\linewidth}
  \caption{Average growth rate of the bubbles generated by the pulsed laser with the 105~µm fiber for a range of pulse energies and 5 different concentrations of ARAC. Error bars indicate standard deviation in pulse energies and bubble growth rate for at least 6 individual bubbles with identical initial conditions. For each concentration, the data is fitted with a linear fit, with corresponding slope (units of 10$^{-9}$~m$^3$/(s*µJ)).}
    \label{fig: Growth_rate_vs_energy_concentrations}
\end{figure}

The bubble growth rate as a function of the pulse energy is plotted in Figure~\ref{fig: Growth_rate_vs_energy_concentrations}, for a range of dye concentrations and the 105~µm fiber tip. For all pulse energies, an increase in dye concentration results in a larger growth rate. This is explained by the fact that an increasing dye concentration increases the absorption coefficient and thus a more localized energy absorption. This results in higher liquid temperatures and more vaporization. 

These results indicate that the bubble and resulting jet dynamics can be controlled by the absorption coefficient of the liquid. This means that a jet injector relying on a pulsed laser is not limited by the range of pulse energies, as an increase or decrease in dye concentration will result in a change in jet velocity and injection depth. Furthermore, by increasing the dye concentration, the required pulse energies will decrease, which would allow the use of lasers to be more affordable and possibly smaller in size.

More specifically, the absorption coefficients we used of 0.7~-~2~$\times$~10$^4$~m$^{-1}$ are similar to the absorption coefficients of water around its peak at 2~µm~\cite{Ruru2012}. Holmium and Thulium lasers operate near this absorption peak and are often used for the irradiation of water and biological tissues~\cite{Wang2020}.  Larger absorption coefficients have been used, either by employing a different dye~\cite{Tagawa2012}, or by using an Er:YAG laser~\cite{Park2012,Jang2013}, where the absorption coefficient of water is approximately two orders of magnitude larger~\cite{Hale1973}. In the first case, even lower pulse energies (E~=~19~µJ) than the ones in this study would result in a bubble and jet (although in a different setup). However, for the second case, the Er:YAG laser has a much longer pulse duration of 250~µs~\cite{Jacques1993,Jang2013}, for which reason there is no thermal or pressure confinement, and it operates similarly to the CW laser. Therefore this laser typically requires pulse energies 400-1000~mJ for bubble and jet formation~\cite{Jang2013,Jang2016}, which is three orders of magnitude larger than used in our study. Besides that, Er:YAG lasers are not compatible with fiber delivery~\cite{Fried2018a}. Near 1450~nm, there is another peak in the absorption coefficient of water, although twice as small ($\alpha$~=~3150~m$^{-1}$~\cite{Ruru2012}) compared to the lower limit in our study. A previous study~\cite{Krizek2021} with a 1574~nm laser ($\alpha\approx$~900~m$^{-1}$) showed that it would require pulse energies of 2-4~mJ, approximately one order of magnitude larger compared to the energies probed here. For even smaller wavelengths, absorption of water is negligible, and thus studies without the use of a dye rely fully on non-linear absorption and require more energy~\cite{Rohilla2020}.

\subsubsection*{Fiber core diameter}
\begin{figure}
\centering
\begin{minipage}{.49\textwidth}
  \centering
  \includegraphics[width=.9\linewidth]{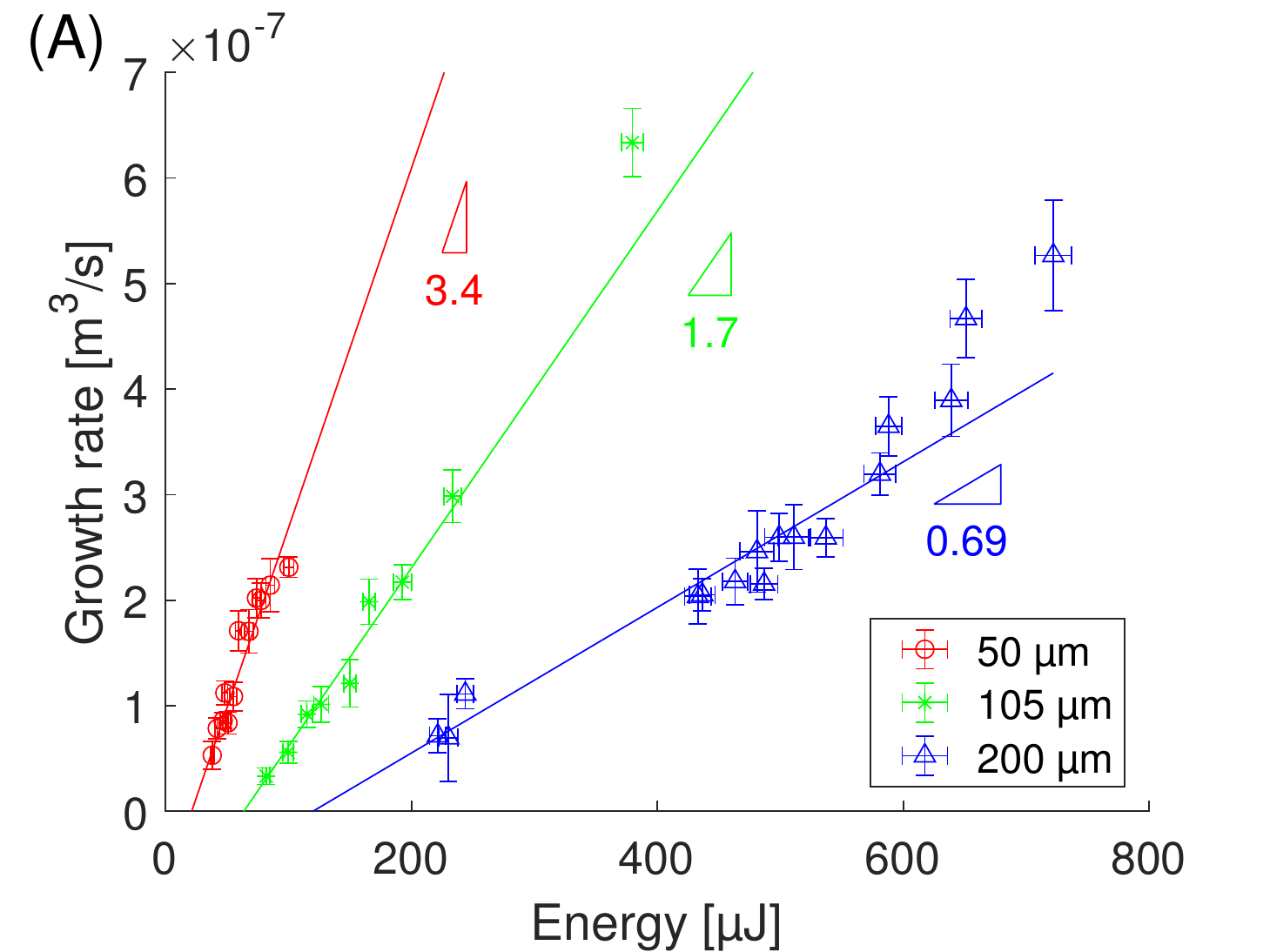}
  \captionsetup{width=.9\linewidth}
  
\end{minipage}
\vline
\begin{minipage}{.49\textwidth}
  \centering
  \includegraphics[width=.9\linewidth]{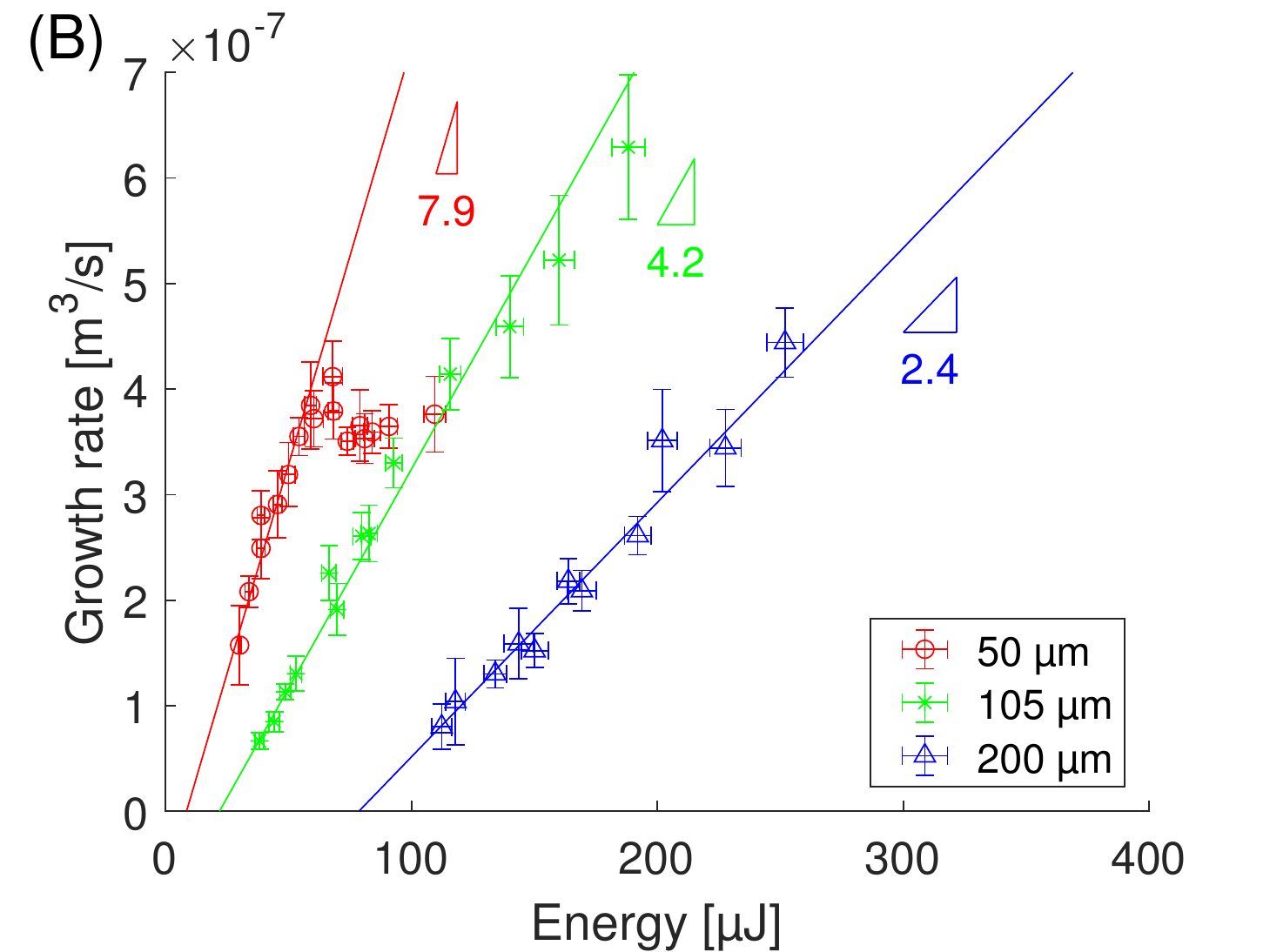}
  \captionsetup{width=.9\linewidth}
\end{minipage}
 \caption{Average volumetric bubble growth rate for three optical fiber sizes. Error bars indicate standard deviation of at least 6 individual measurements. Solid lines indicate the best linear fit, with corresponding slope (units of 10$^{-9}$~m$^3$/(s*µJ)). (A): Dye concentration of C$_\textnormal{dye}$~=~10~mM. (B): Dye concentrations of C$_\textnormal{dye}$~=~52~mM. }
 \label{fig: Fiber radius 10 and 52 mM}
\end{figure}

Figure~\ref{fig: Fiber radius 10 and 52 mM} shows the volumetric bubble growth rates for the three different fiber core diameters: 50, 105 and 200~µm. Figure~\ref{fig: Fiber radius 10 and 52 mM}A is for the ARAC concentration of 10 mM and Figure~\ref{fig: Fiber radius 10 and 52 mM}B is for the ARAC concentration of 52 mM. The measurement data per fiber are fitted with the best linear fit. For both concentrations, larger fibers require larger pulse energy to obtain the same growth rate. This trend was also observed for jet velocity and fiber sizes~\cite{Krizek2020}, and explained by a larger energy density for the smaller fibers, and thus a larger pressure. This means that a smaller fiber with a less powerful laser can create the same bubble and resulting jet.

However, for the 50~µm fiber, we could not reach very large growth rates for two reasons: First, for the 52~mM dye concentration, we observed a plateau in the growth rate for pulse energies larger than 60~µJ. Second, the tip of the 50~µm fiber is very prone to damage - because of high light power densities, and the laser-induced damage threshold was found at approximately 120~µJ. This means that the larger fibers are more flexible in their use, although they require larger pulse energies.

\subsubsection*{Energy threshold for bubble nucleation}
\begin{figure}
    \centering
    \includegraphics[width = 0.7\linewidth]{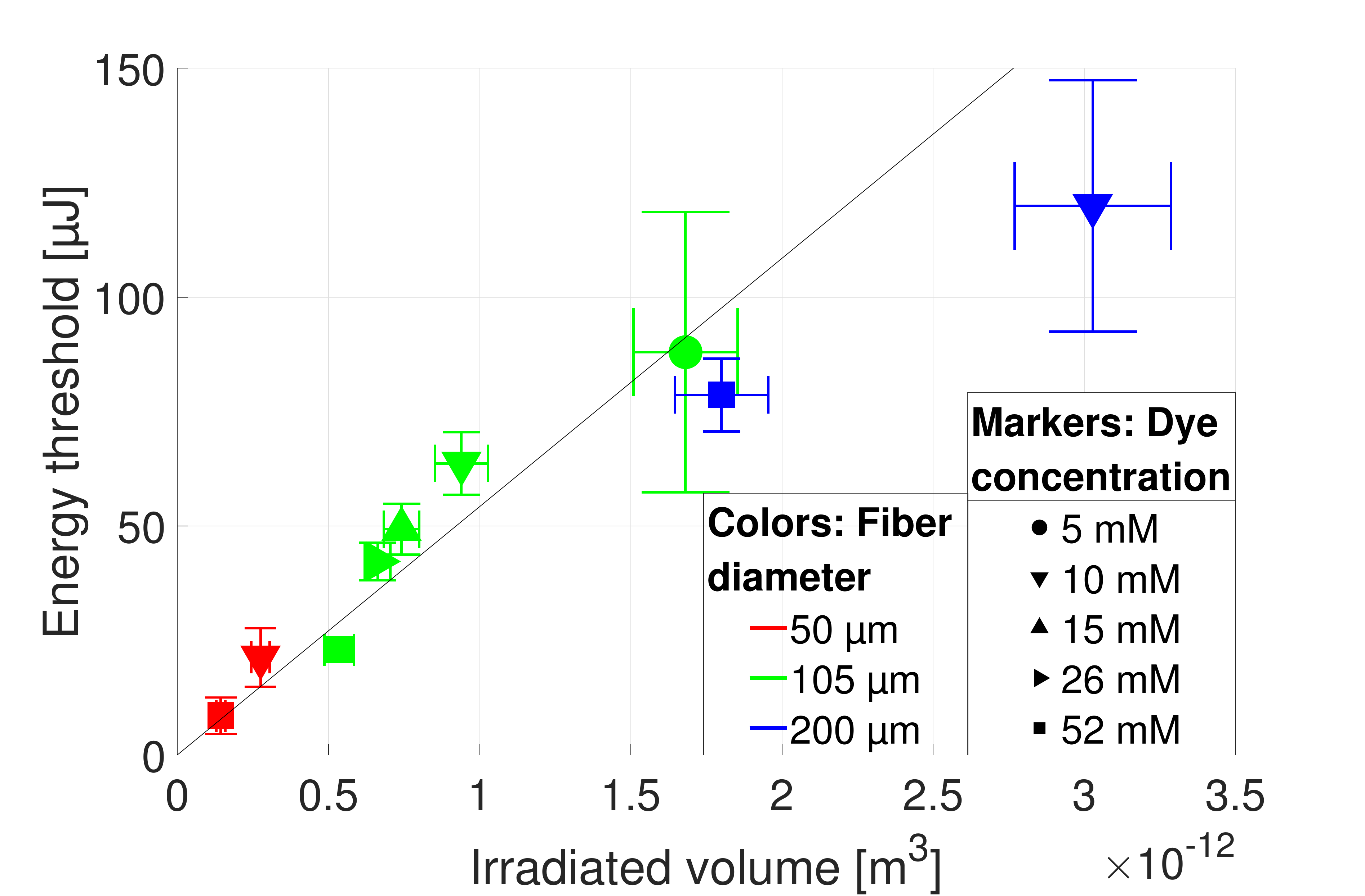}
    \caption{Energy threshold taken from offset in the linear fits from Figures~\ref{fig: Growth_rate_vs_energy_concentrations}~and~\ref{fig: Fiber radius 10 and 52 mM}, plotted against the irradiated volume. Colors indicate the different fiber diameters, symbols indicate dye concentrations and error bars indicate 95\% confidence interval for the linear fits. The black line indicates the best linear fit.}
    \label{fig: threshold_vs_volume_v2}
\end{figure}

The intersection with the x-axis of the linear fits in Figures~\ref{fig: Growth_rate_vs_energy_concentrations}~and~\ref{fig: Fiber radius 10 and 52 mM}, indicates an energy threshold for bubble formation. These energy thresholds are plotted against the irradiated volumes in Figure~\ref{fig: threshold_vs_volume_v2}. The irradiated volume is calculated until the typical absorption length $\delta$, assuming a conical shape (due to divergence) starting at the fiber tip with a diameter equal to the core diameter (see Appendix B). The linear trend of the energy threshold with the irradiated volume indicates a constant energy density required for bubble nucleation. Therefore, the energy threshold scales quadratic with the fiber radius and linear with the absorption length. However, in the case where the absorption length is much larger compared to the fiber radius, the beam divergence should be taken into account. In practical applications where less powerful lasers are preferred, the fiber should have a small core radius, small NA (small divergence) and the liquid should have a large absorption coefficient.

\subsection{CW laser}\label{sec: CW laser}
\subsubsection*{Nucleation time and delivered energy}
\begin{figure}[t!]
    \centering
\begin{minipage}{.49\textwidth}
  \centering
  \includegraphics[width=.9\linewidth]{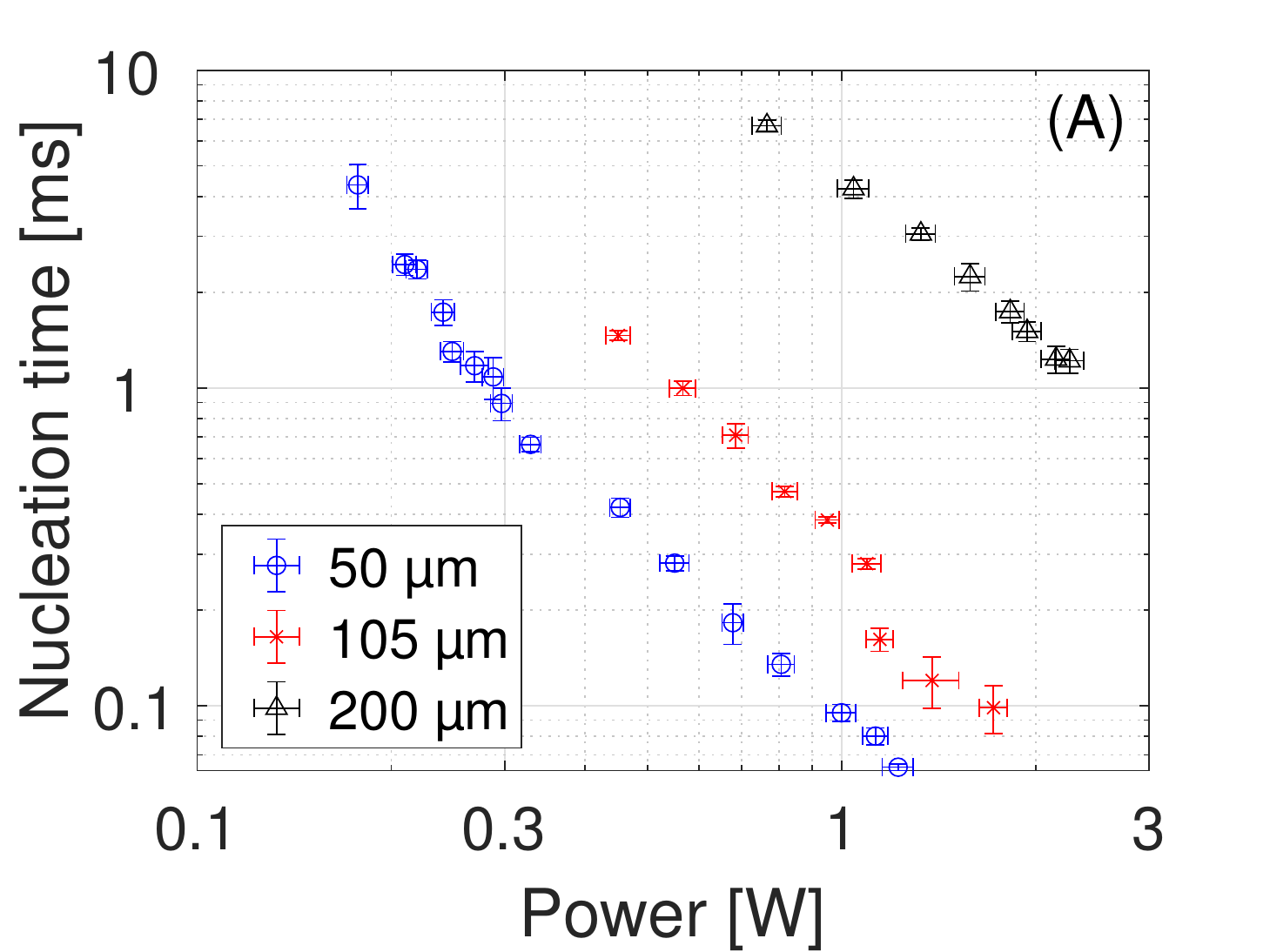}

\end{minipage}
\vline
\begin{minipage}{.49\textwidth}
  \centering
  \includegraphics[width=.9\linewidth]{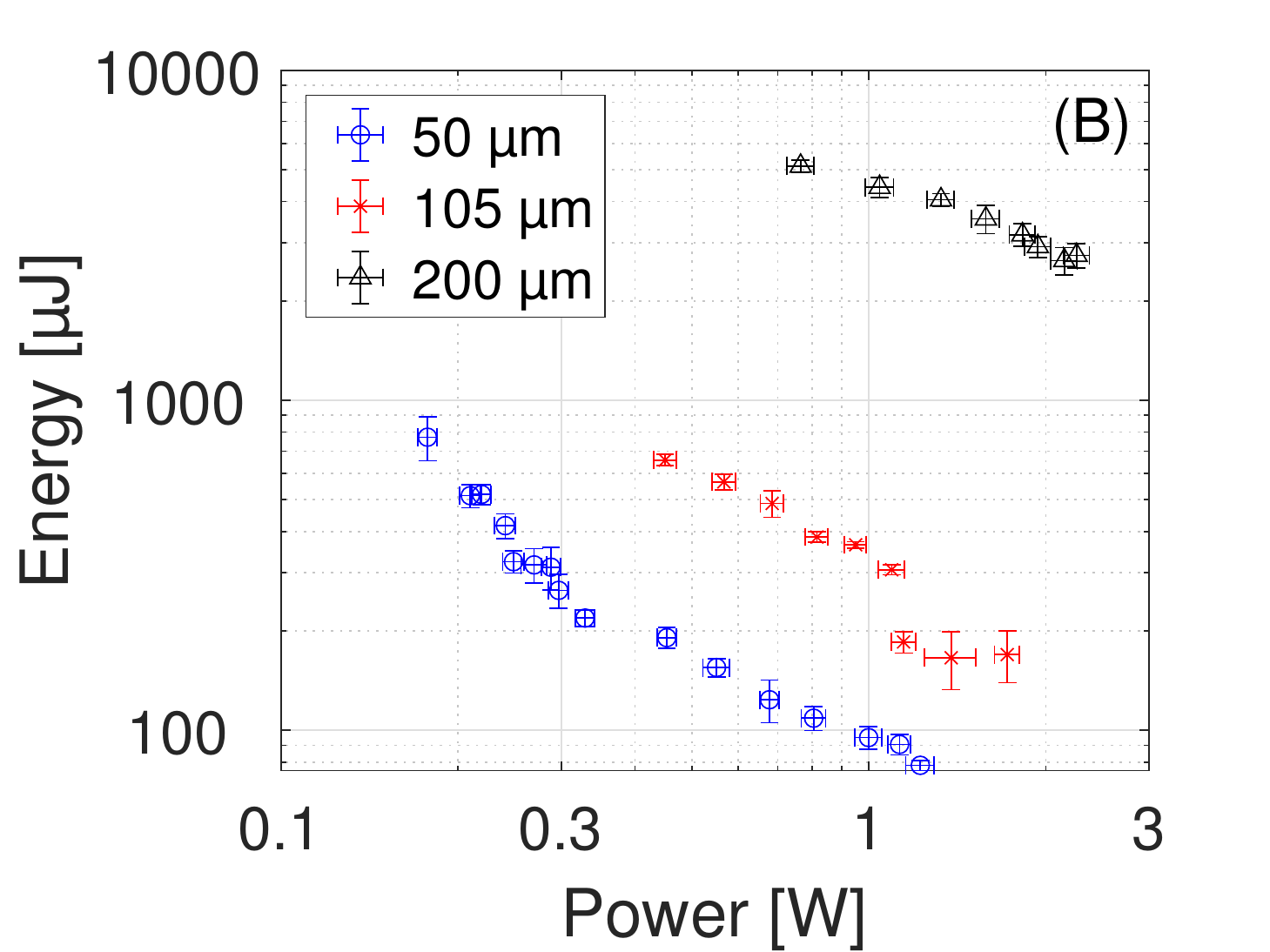}

\end{minipage}
    \caption{Nucleation time (A) and delivered energy (B) as a function of CW laser power for three optical fibers with different diameter. Each data point is an average over at least 6 individual measurements, with the error bars indicating the standard deviation.}
    \label{fig: CW power nucl time energy}
\end{figure}

In contrast to the pulsed laser, for the CW experiment, only the power can be directly controlled, and not the energy. Here, the delivered energy depends on the power and the moment of nucleation, which depends on the laser power~\cite{Padilla-Martinez2014}. Figure~\ref{fig: CW power nucl time energy}A shows the influence of the laser power on the nucleation time for the three fibers. First, the nucleation time increases with decreasing laser power, as it takes longer to reach the nucleation temperature. Second, for fixed laser power, the nucleation time increases with increasing fiber diameter. This is explained by an increase in beam size and thus a decrease in intensity, for which reason it takes longer to reach the nucleation temperature. Due to a limited range of laser power, sub-ms nucleation times could not be reached for the 200~µm fiber. 

Besides the influence of the power and fiber size, there are fluctuations in nucleation time for each individual data point. This fluctuation in nucleation time is called \textit{jitter}~\cite{Padilla-Martinez2014} and is explained by the stochastic nature of nucleation. These fluctuations linearly affect the delivered energy, and thus reduce the reproducibility. We find that the typical fluctuation (standard deviation) is approximately 8\% of the nucleation time, which is smaller compared to earlier findings up to approximately 60\%~\cite{Padilla-Martinez2014,Zhang2022}. We attribute this to the fact that bubbles develop directly at the optical fiber, instead of heating a liquid with a non-collimated laser beam, where the beam size is more difficult to control and reproduce. Furthermore, in our case we do only create individual bubbles, after which the liquid cools down again, which may result in a more reproducible nucleation time. 

The delivered energies, which are calculated by multiplying the nucleation time and the power, are shown in Figure~\ref{fig: CW power nucl time energy}B. Similar to the nucleation times, a reduction in power or an increase in fiber size results in an increase of the energy delivered. This is because an increase in nucleation time results in an increase in heat diffusion. This heat diffusion will result in an even longer nucleation time as the local temperature increases at the fiber tip is slower, which then results in larger delivered energy. 
For the 50 and 105~µm fibers, the CW laser results in comparable energies as the ns-laser.
However, for CW the energies required for bubble formation by the 200~µm fiber are an order of magnitude larger due to the longer nucleation times, which is caused by the larger fiber area and thus lower intensities. 

Previous work of Ref.~\cite{OyarteGalvez2020} refers to a free-space laser diode with free-space light focusing with a 6*33~µm$^2$ laser beam. For a power of 0.5~W, they found a nucleation time of 600~µs, resulting in a delivered energy of 300~µJ, for a liquid with a similar absorption coefficient as ours~($\alpha$~=~1.0~and~1.2~*~10$^4$~m$^{-1}$ respectively). The beam size for the smallest fiber in our study is already 10 times larger and still results in twice as fast nucleation time for 0.5~W ($\sim$~300~µs). Therefore, the previous study with free-space optics and a glass microdevice showed a significant optical loss between the laser diode and the microfluidic channel. Here, due to the direct contact of the fiber and the liquid, the number of interfaces is minimized and can be no losses due to misalignment, resulting in higher energy transfer efficiency. 

\subsubsection*{Bubble growth rate}
\begin{figure}[t!]
    \centering
\begin{minipage}{.49\textwidth}
  \centering
  \includegraphics[width=.9\linewidth]{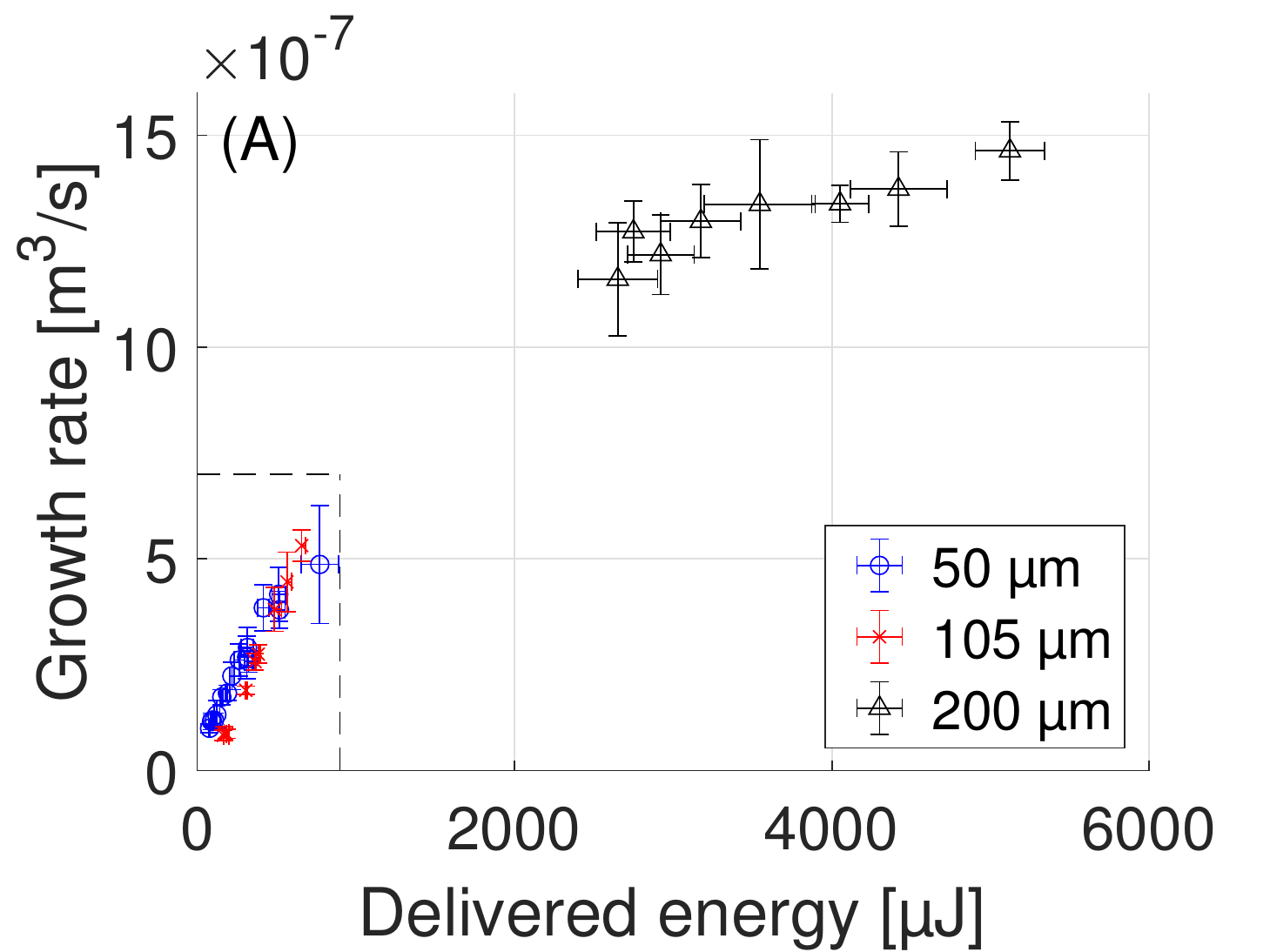}

\end{minipage}
\vline
\begin{minipage}{.49\textwidth}
  \centering
  \includegraphics[width=.9\linewidth]{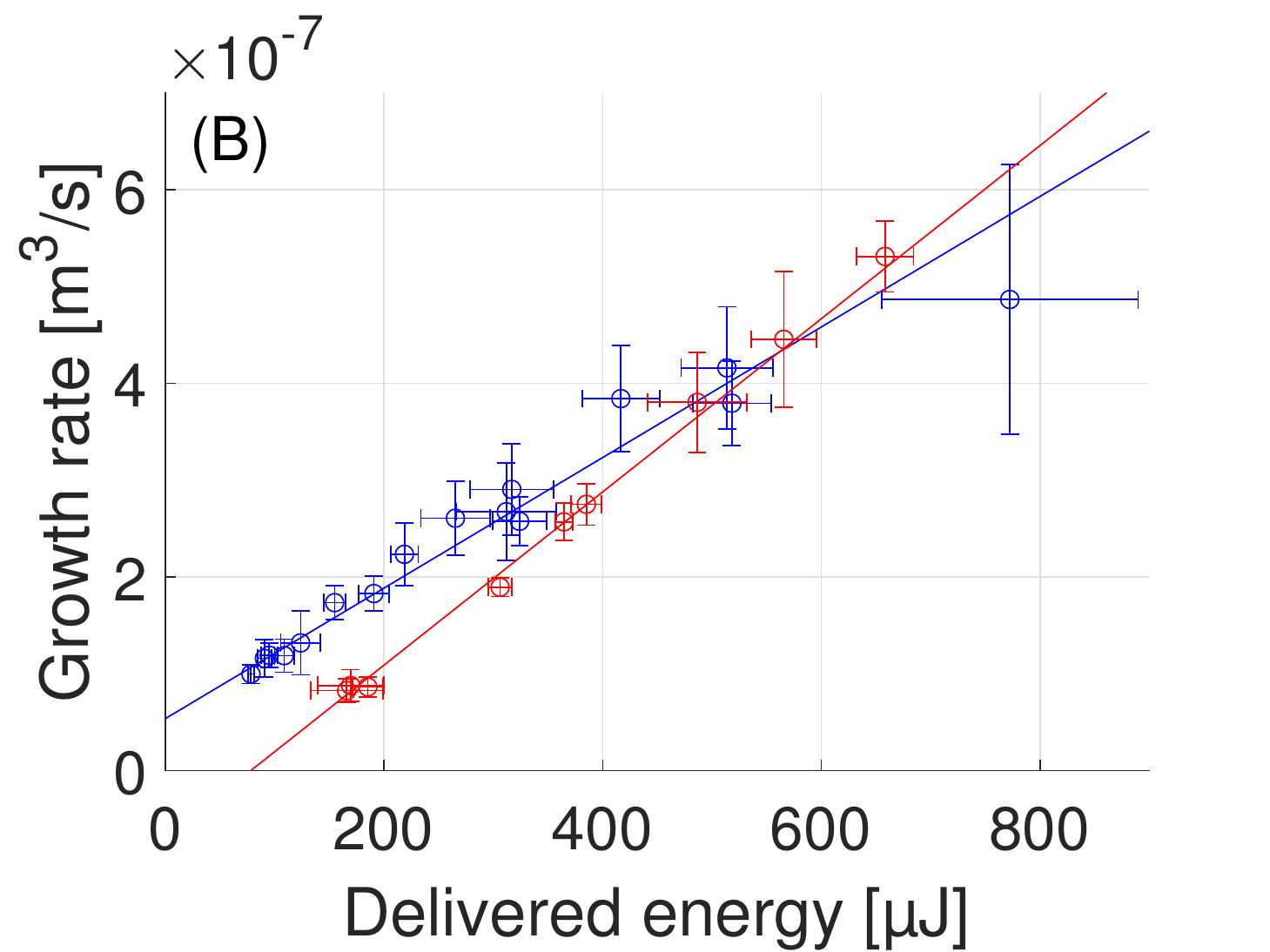}

\end{minipage}
    \caption{Average volumetric bubble growth rate for the CW laser. Figure (A) includes all three fiber sizes. As the energy and growth rates for the 200~µm fiber are much larger compared to the 50 and 105~µm fiber, figure (B) shows a close-up of the left bottom of (A), indicated by the dotted lines. Figure (B) includes the best linear fit for the 50 and 105~µm fiber.}
    \label{fig: CW growth rates fibers}
\end{figure}

The bubble growth rates for the CW laser source are shown in Figure~\ref{fig: CW growth rates fibers}A. It is clear that the growth rates of the bubbles generated with the 200~µm fiber are much larger compared to the 50 and 105~µm fiber. The increased growth rate is explained by the delivered energy, which is an order of magnitude larger compared to the other fibers. As discussed above, smaller energies could not be delivered with the used CW laser for this fiber as it would require much larger powers. The bubbles generated at the 200~µm fiber are also much larger compared to the other two fibers, and all generated bubbles grow beyond the edge of the capillary, for which reason they coalesce with the surrounding air before reaching their maximum volume. An example of such a bubble generated at the 200~µm fiber is shown in Appendix C.). 

Figure~\ref{fig: CW growth rates fibers}B shows a close-up of the growth rates of the bubbles generated at the two smaller fibers. It shows that for the CW laser, the bubble growth rate also increases linearly with increasing delivered energy. For most energies, the 50~µm fiber has a larger growth rate compared to the 105~µm fiber, which can be explained by the larger energy density. However, for delivered energies larger than 500~µJ, the 105~µm fiber results in faster-growing bubbles. We attribute it to the longer nucleation time for the 50~µm fiber to reach those levels of energy. For the 50~µm fiber, the nucleation times to reach E~$>$~500~µJ is t$_\textnormal{n}~>$~2ms, whereas for the 105~µm fiber, the nucleation time is only half. Therefore there is increased heat dissipation for the 50~µm fiber, and thus a smaller efficiency. Therefore, the growth rate no longer increases linearly with the delivered energy, and larger growth rates require much more energy for the 50~µm fiber. 

Interestingly, extrapolating the linear fit of the 50~µm fiber in Figure~\ref{fig: CW growth rates fibers}B shows a positive growth rate for zero energy, which is physically not possible. However, due to thermal dissipation, it is questionable whether the growth rate for the CW-laser would actually be linear with the energy, as thermal dissipation may reduce the energy efficiency. Therefore, even though the initial behaviour for both fibers seems linear, extrapolating this to find an energy threshold does not seem to have a physical meaning.

\subsubsection*{Secondary bubbles}
\begin{figure}[t!]
\centering
\begin{minipage}{.45\textwidth}
  \centering
  \hspace{-1cm}
  \includegraphics[width=1.05\linewidth]{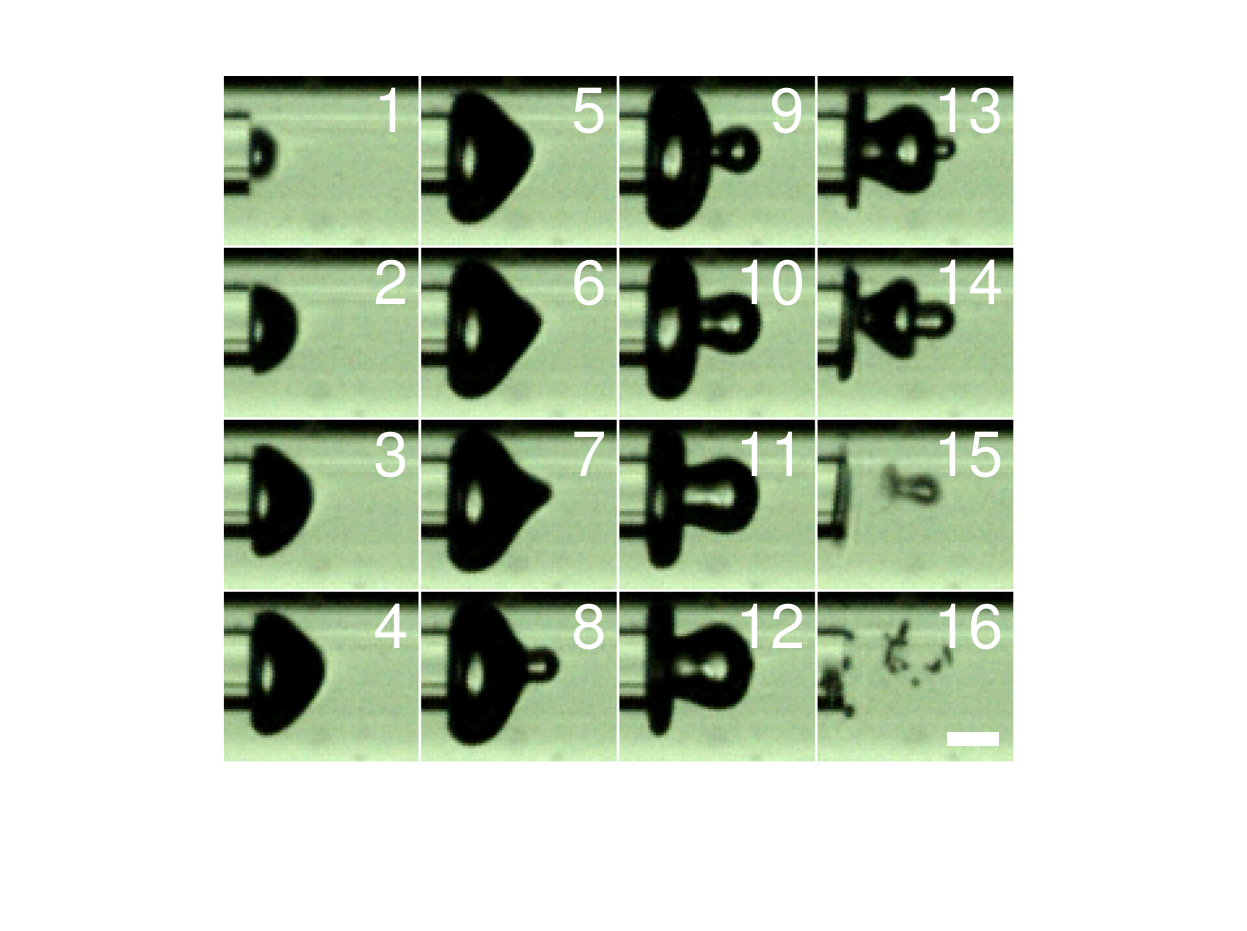}

\end{minipage}
\vline
\begin{minipage}{.45\textwidth}
  \centering
  \includegraphics[width=.9\linewidth]{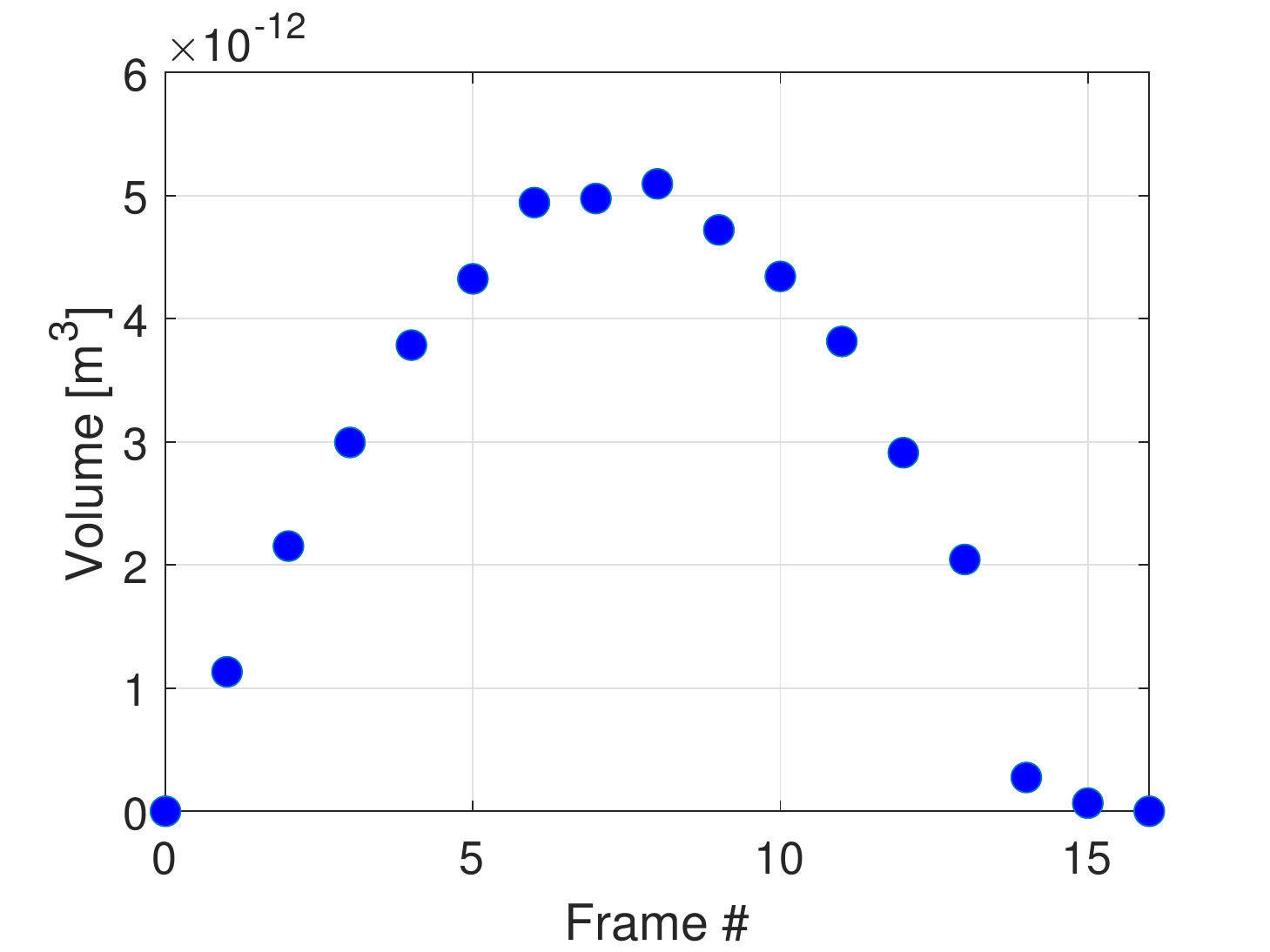}
  
\end{minipage}
    \caption{\textbf{Left:} 16 consecutive frames showing the typical dynamics of a small bubble generated by the CW laser. As the laser is still running during the bubble growth, more liquid will be heated, which results in a secondary (frame 7-12) and even a third bubble (frame 13-14). Snapshots are consecutive, from top to bottom and left to right, with an inter-frame time of 5~ms. The initial nucleation time (top left frame) was 135~µs, and the bubble reached its maximum size (frame 8) at 175~µs. Scalebar on the right bottom measures 100~µm. \textbf{Right:} Calculated bubble volume for each frame, showing that the initial maximum is reached at frame 6, but due to secondary bubble formation, another maximum is reached at frame 8, although the difference is small.}
    \label{fig: Secondary bubble}
\end{figure}

For small bubbles generated by the CW laser (R~$<$~150~µm), we observe the formation of a secondary bubble at the vapor-liquid interface, see~Figure~\ref{fig: Secondary bubble}. These events are different from twin cavitation bubbles reported in Ref.~\cite{Zhang2022}, where the initial bubble first partially collapses before forming a secondary bubble. In our case, the secondary bubble typically forms when the initial bubble reaches its maximum. Furthermore, in our case the volume of the secondary bubble is smaller compared to the first one, for which reason its appearance does not affect the parabolic shape of the bubble volume versus time.
These secondary bubbles can be explained by the fact that the laser remains irradiating during the bubble formation and, therefore, further heats the liquid during the bubble growth. In the case of large laser powers and short nucleation times, this heating during the bubble lifetime could create a secondary bubble. In the case of Figure~\ref{fig: Secondary bubble}, the growth time of the bubble takes 40~µs, whereas the initial nucleation time was 135~µs, which means that after nucleation, another 30\% of the initial energy is still delivered, which results in secondary nucleation at the vapor-liquid interface of the bubble. 

The occurrence of these secondary bubbles depends on the laser power and nucleation time of the initial bubble. The secondary bubble forms during the lifetime of the bubble (typically 50-200~µs), during which enough additional energy has to be delivered to form a secondary bubble. We find that these secondary bubbles only appear when the intensities are sufficiently larger such that the initial nucleation time is smaller than approximately 450~µs, both for the 50 and 105~µm fiber. For the 200~µm fiber the intensities are smaller and thus the nucleation times are much larger we did not observe any secondary bubbles. However, we hypothesize that the use of a more powerful laser can also form these secondary bubbles with the 200~µm fiber. However, as mentioned, these secondary bubbles only happen for small nucleation times and large powers, which result in small bubbles overall. Therefore, we conclude that the larger and faster-growing primary bubbles are of more interest.

\subsection{Comparison between pulsed and CW}

A direct comparison between the bubbles generated by the pulsed and CW laser requires a matching absorption coefficient such that the volumes over which the energy is absorbed are identical. Therefore, we compare the results of the 10~mM ARAC solution, as its absorption coefficient at 532~nm is nearly identical compared to the absorption coefficient of water at 1950~nm. In this subsection, we compare the bubbles generated by the 50 and 105~µm fiber, as they show similar results in the growth rate (1~-~$7*10^{-7} \textnormal{m}^3/s$). The 200~µm fiber shows very different results for the CW laser, as the delivered energies are much larger (see Figure \ref{fig: CW growth rates fibers}). 

\subsubsection*{Energy efficiency}
\begin{figure}[t!]
    \centering
    \includegraphics[width=0.5\textwidth]{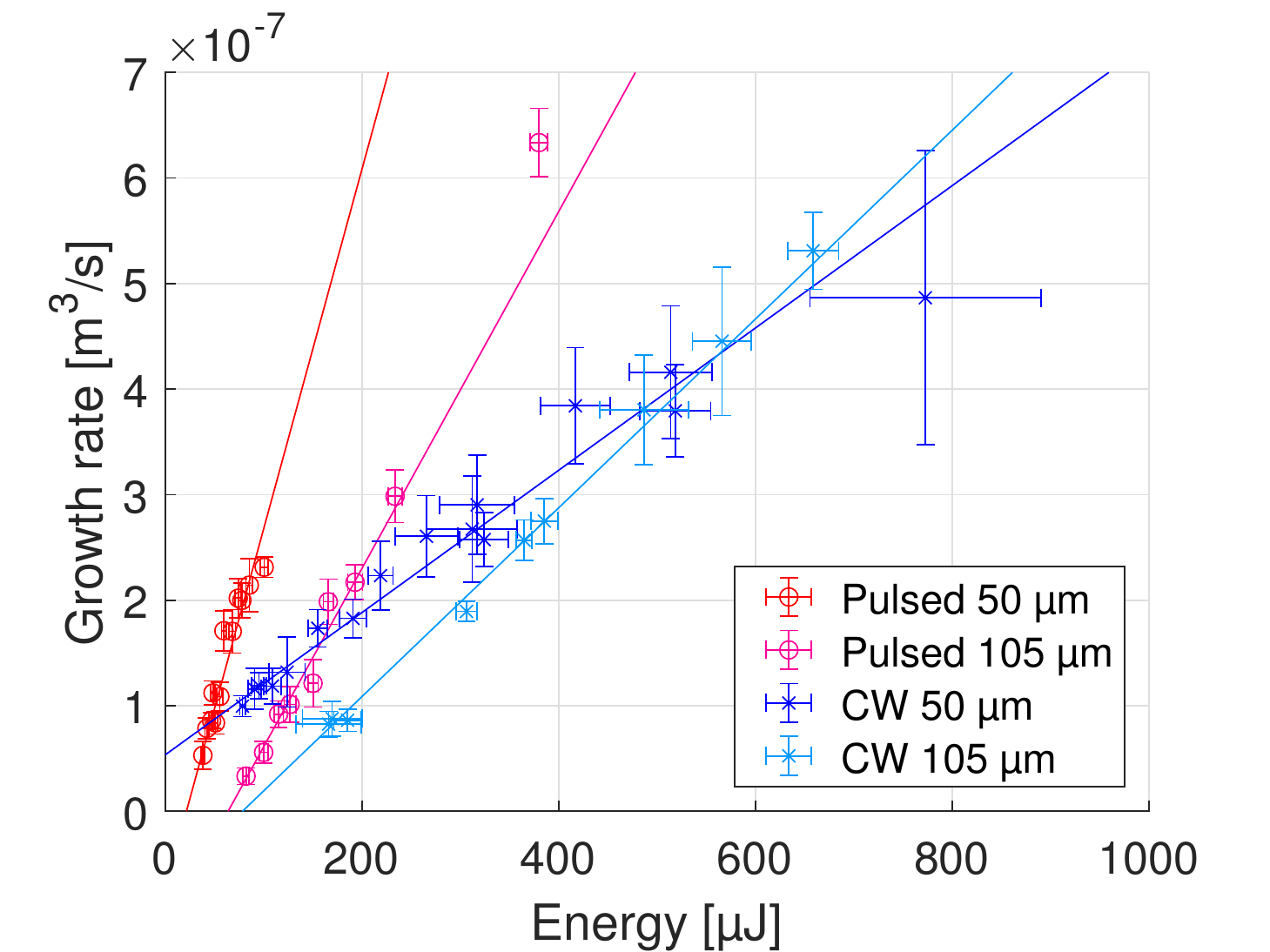}
    \captionsetup{width=0.5\linewidth}
    \caption{Growth rate of bubbles generated by the pulsed (red) and CW laser (blue), for fibers with core diameters of 50 (dark colors) and 105~µm (light colors). The absorption coefficients for both laser are the same ($\alpha$~$\approx$~12000~m$^{-1}$).}
    \label{fig: GR_pulsed_CW_50_105}
\end{figure}

Budgeting of energy has practical implications on the right choice of laser source.
The growth rates for the CW and pulsed laser are shown in Figure~\ref{fig: GR_pulsed_CW_50_105}, for the 50 and 105~µm fiber. For the same delivered optical energy, the pulsed laser results in faster growth rates, although the required energies are in the same order of magnitude. Typically, the CW laser requires two to three times more optical energy than the pulsed laser to generate a bubble with the same growth rate. The reduced efficiency is explained by three reasons. First of all, for the bubbles generated by the CW laser heat diffusion cannot be neglected, as the nucleation time is 0.1~-~5~ms, which is in the same order as the thermal diffusion timescale ($\sim$~4~ms, see Equation~\ref{eq: thermal diffusion}). Most of this heat will be lost, as it will heat up the liquid around the irradiated volume. Though, this heat increase in the proximate area would be limited, and most of it does not result in the phase transition. Second, the absorption coefficient of water around 1950~nm decreases with increasing temperature, and the absorption coefficient at 100$\degree$C is only half the initial value at room temperature~\cite{Jansen1994,Lange2002}. Therefore, for the heating phase with the CW laser, the average absorption length is much larger compared to the absorption length of the pulsed laser experiments.
Third, for the pulsed laser, there might be non-linear absorption due to the high peak power. The measured absorption coefficient of the 10~mM ARAC is the same as the absorption coefficient of water for low optical intensities, but the absorption coefficient for the high-power pulses may be larger due to additional non-linear absorption by the liquid. This increased absorption would result in a smaller irradiated volume and thus a higher energy density, which will increase the bubble growth rate.

\subsubsection*{Bubble dynamics}
\begin{figure}
\centering

\begin{minipage}{.49\textwidth}
  \centering
    \includegraphics[width=0.9\textwidth]{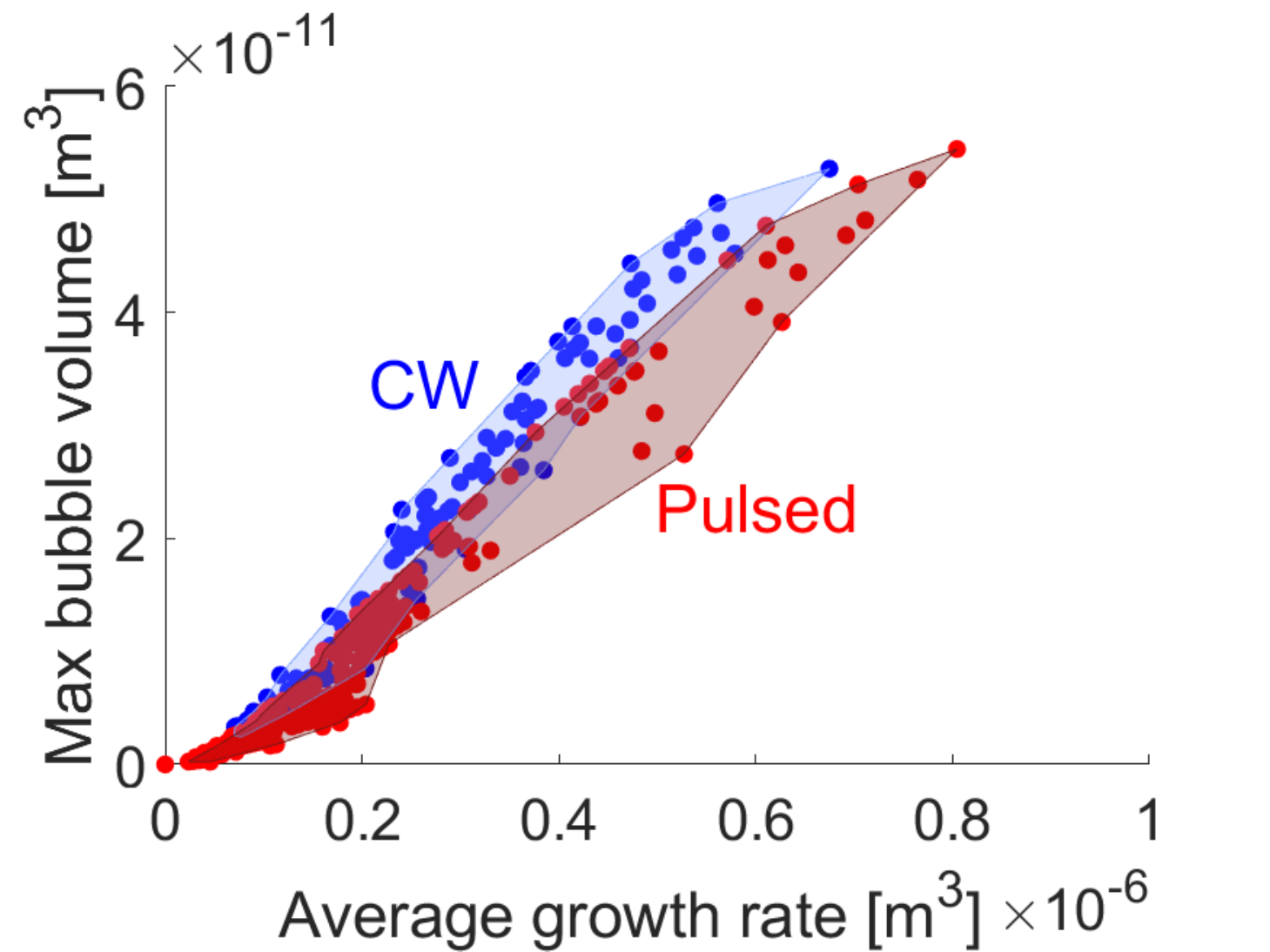}
    \captionsetup{width=0.9\linewidth}
    \caption{Maximum bubble volume vs average growth rate for individual bubbles generated by the pulsed laser (red) and the CW laser (blue). The shaded area encapsulates all data.}
    \label{fig: GR_MV_pulsed_CW}
\end{minipage}
\vline
\begin{minipage}{.49\textwidth}
  \centering
    \includegraphics[width=0.9\textwidth]{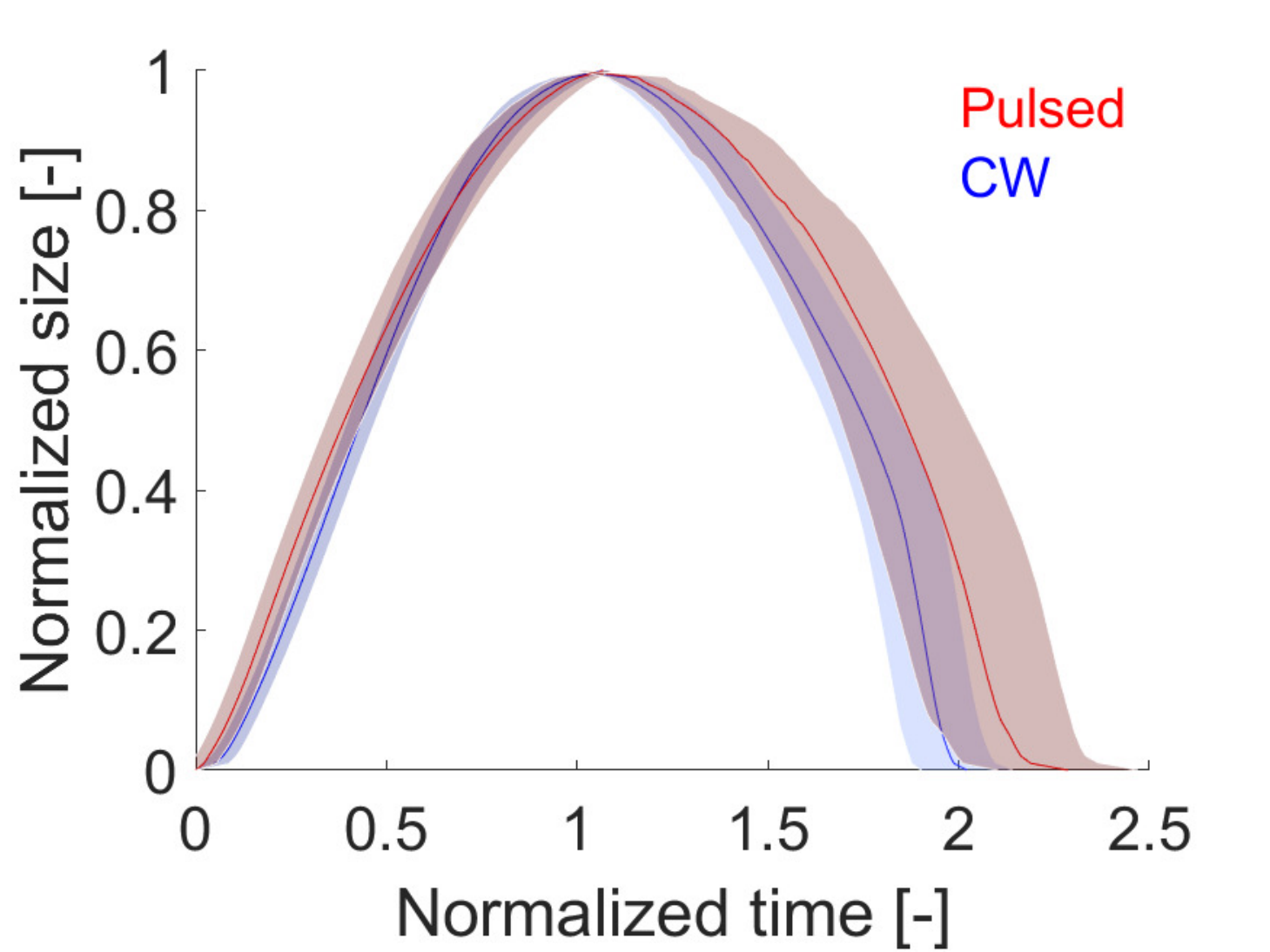}
    \captionsetup{width=0.9\linewidth}
    \caption{Normalized bubble size vs normalized time, for CW (blue) and pulsed (red). Lines indicate averages over all bubbles, shaded regions indicate standard deviation. For each individual bubble, the volume is normalized by the maximum volume, and the time is normalized by the growth time.}
    \label{fig: Volume_time_normalized}
\end{minipage}
\end{figure}

Bubble dynamics govern the subsequent microfluidic jet parameters.
As shown above, the typical growth rates are similar for both lasers. However, identical average growth rates may not result in the exact same bubble characteristics. An increase in growth time, results in a larger bubble, assuming the average growth rate is the same. The maximum volume against the average growth rate is shown in Figure~\ref{fig: GR_MV_pulsed_CW} for the bubbles generated by the pulsed (red) and CW (blue) laser. For identical average growth rates, the bubbles created by CW laser grow to a $\sim 15\%$ larger volume (due to the increased growth time). Alternatively, for bubbles with the same maximum volume, the bubbles created by the CW laser take longer to reach that volume. This reduced growth rate could be explained by the larger volume over which the delivered energy is dissipated, resulting in a less explosive phase transition. However, this also means that for the same average growth rate, the bubbles generated by the CW laser will push out the remaining liquid over a longer length and time. This increased time of energy transfer will mostly affect the jet tail, which is typically slower and more dispersed~\cite{Krizek2020,AndrewUnpublished}. We hypothesize that this longer growth time results in a faster and more reproducible jet tail.

A comparison between the growth and collapse can be made by normalizing the volume and time, as shown in Figure~\ref{fig: Volume_time_normalized}. For each individual bubble, the volume is normalized by its maximum volume and the time by the growth time. The lines show an average value over all bubbles, and the shaded region is the standard deviation. This figure shows that the initial normalized growth rate of the bubbles generated by the pulsed laser is larger compared to bubbles generated by the CW laser, as indicated by the steeper curve for small times. 
Furthermore, it shows that for the pulsed laser, the collapse is on average $\sim 20\%$ slower compared to the growth, whereas for the CW they take equal time. Both of these findings are in agreement with previous studies on the bubble dynamics for pulsed laser~\cite{Zwaan2007,Sun2009} and CW~\cite{OyarteGalvez2020}. However, this difference is mainly caused by the $\sim 15\%$ smaller growth time for the pulsed laser, see Figure~\ref{fig: GR_MV_pulsed_CW}. If both curves were normalized by the same time, the time of collapse would only vary by $<5\%$.

\subsubsection*{Reproducibility}
Reproducibility is an important factor considering application in health care. The error bars in Figure~\ref{fig: GR_pulsed_CW_50_105} show that the deviation in the bubble growth rate is larger for the CW laser (blue colors) compared to the pulsed laser (red colors). This indicates that even with identical initial conditions for the CW laser, the bubble dynamics can vary each time slightly. As discussed in section~\ref{sec: CW laser}, the energy delivered by the CW laser is not controlled directly but depends on the laser power and nucleation time. As nucleation is a stochastic event, there is a variance in the nucleation time, resulting in less or more delivered energy for early or late nucleation, respectively. 

\begin{figure}[t!]
    \centering
    \includegraphics[width=1\textwidth]{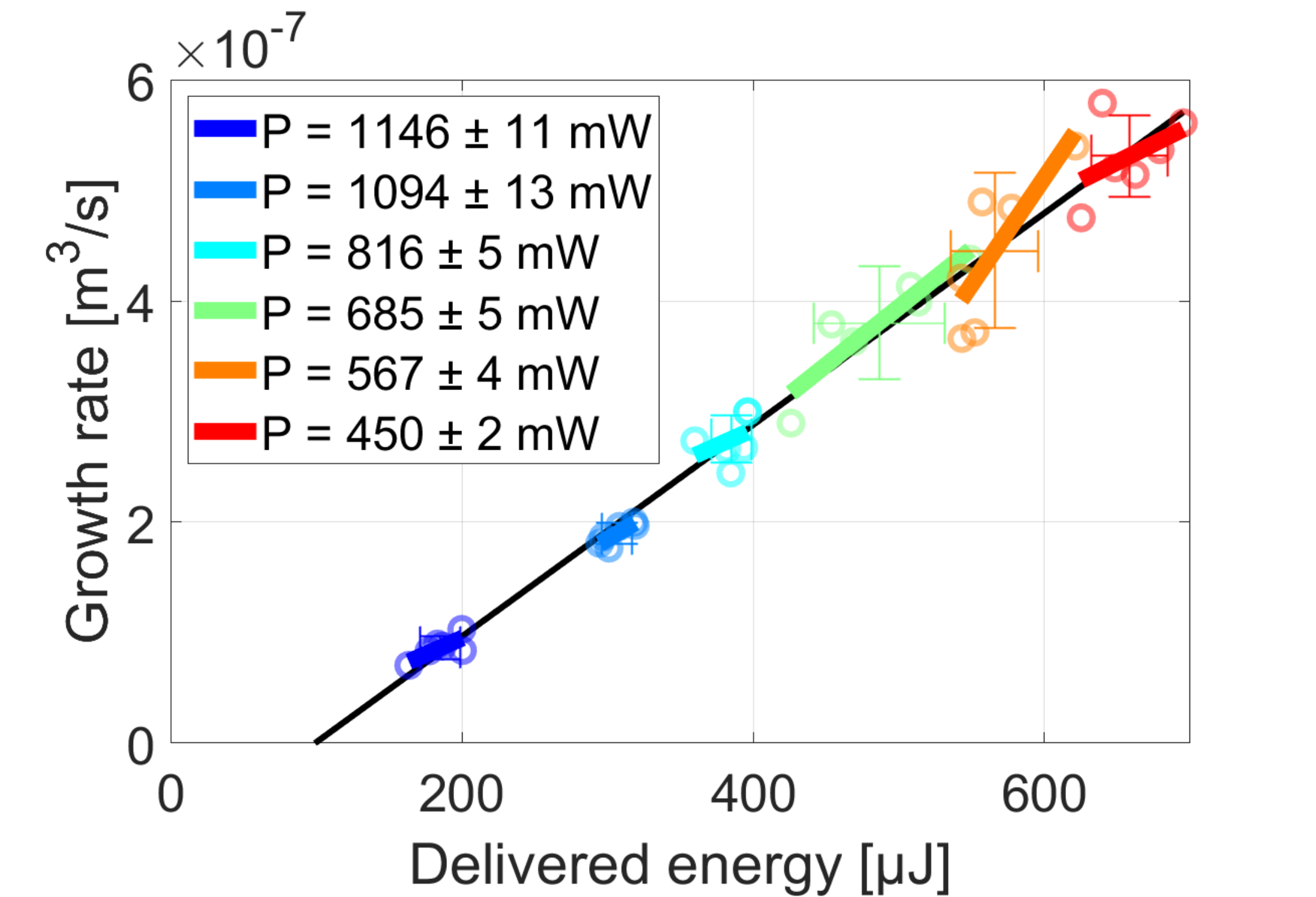}
    \captionsetup{width=1\linewidth}
    \caption{Growth rate for individual bubble events for the 105~µm fiber with the CW laser. Each color indicates a different laser power. The delivered energy mainly depends on the laser power but also on the moment of nucleation (e.g. late nucleation results in larger delivered energy). The circles are individual measurements, and the error bars indicate the mean and standard deviation of those individual measurements with the same laser power (at least 6). The colored lines indicate the best linear fit per laser power. The black line indicates the best linear fit for all data points. The slope of the colored lines is similar to the black line, indicating a correlation between the time of nucleation and bubble growth rate.}
    \label{fig: CW growth individual}
\end{figure}

Figure~\ref{fig: CW growth individual} shows the growth rate of all individual bubbles generated by the CW laser together with the mean and standard deviation for a group of 6 or more with identical initial conditions (laser power). Each color indicates a different laser power, which influences the delivered energy. However, even for measurements with identical laser power, the delivered energy can vary up to $\pm~50$~µJ. For each of these laser powers, the individual results are typically in the bottom-left or top-right quadrant of the error bars. This indicates that there is a correlation between the additional energy for delayed nucleation and the bubble growth rate. Furthermore, per laser power, the individual data points are fitted with a linear fit, which is shown by the colored lines. The slope of these colored lines is very similar to the slope of the black line, which is the best fit when comprising all data. Therefore, the variance in nucleation time directly affects the bubble growth rate, which means that delayed nucleation results in more energy and a faster-growing bubble, and early nucleation results in less energy and a slower-growing bubble. However, as nucleation is a stochastic process, this results in a random deviation and reduces reproducibility.

For the pulsed laser, the standard deviation in delivered energy is only affected by the laser specifications. On average, it is much smaller with 2\% of the delivered energy, compared to 8\% for CW. As this deviation also directly affects the bubble growth rate, the bubble growth rate for the pulsed laser is thus more reproducible. 

\subsection{Practical considerations of laser choice for applications}
Engineering a medical device based on laser-induced cavitation is limited by the available offer of laser sources on the market. Yet, choosing the right laser emitter is vital for the success of any potential product or prototype, where not only the best physical and optical parameters shall be considered. Other aspects like cost, the robustness of operation, size or compliance with safety hazards play a key role. Intuitively, the cheaper, smaller and safer source is favourable, but these are usually contradictory parameters. The results of this study bring a better understanding of what kind of laser parameters are critical to generate fast traveling liquid droplets of relevance for its use in eventual devices using laser-induced cavitation, e.g. needle-free jet injectors. 
Although an in-depth techno-economical discussion on the selection of laser sources for specific applications is beyond the scope of this study, we will briefly discuss the impact of the different laser and device parameters here. The main differences are shown in Table \ref{Table: CW vs Pulsed}.

For applications where reproducibility and reliability of the bubble dynamics are most important, the pulsed laser is preferred over the CW laser. Due to the reduced control over the delivered energy by the CW laser, there is a larger deviation in individual bubble growth rates. In the case of a jet injection device, this affects the jet velocity and thus injection depth. However, the exact influence of the deviation in bubble dynamics on the injection depth should be investigated in a further study.

\begin{table}[t]
\caption{Overview of the difference of the laser and bubble characteristics of the CW and pulsed laser.}
\label{Table: CW vs Pulsed}
\begin{tabular}{|l|l|l|}
\hline
 & \textbf{Pulsed laser} & \textbf{CW laser} \\ \hline
\textbf{Timescale} & 5~ns & 1~ms \\ \hline
\textbf{Peak power} & 10$^4$~W & 1~W \\ \hline
\textbf{Typical price$^*$} & $\sim$~1000~-~50000~\$ & $\sim$~100~-~1000~\$ \\ \hline
\textbf{Typical size} & Table top device & Handheld device \\ \hline
\textbf{Control over delivered energy} & Direct (input of laser) & Indirect (depends on P \& t$_\textnormal{n}$) \\ \hline
\textbf{Reproducibility (variation)} & 2\% & 8\% \\ \hline
\textbf{Ratio growth to collapse time} & 1:1.2 & 1:1 \\ \hline
\textbf{\begin{tabular}[c]{@{}l@{}}Influence of increase in fiber \\ size (laser input parameters \\ constant)\end{tabular}} & \begin{tabular}[c]{@{}l@{}}Reduction in energy density\\ Increase in energy threshold\\ Smaller/slower growing bubbles\end{tabular} & \begin{tabular}[c]{@{}l@{}}Delayed nucleation\\ Increase in delivered energy\\ Larger/faster growing bubbles\end{tabular} \\ \hline
\textbf{\begin{tabular}[c]{@{}l@{}}Influence of increase in \\ absorption coefficient/\\ dye concentration\end{tabular}} & \begin{tabular}[c]{@{}l@{}}Larger energy density\\ Decrease in energy threshold\\ Increase of bubble size\end{tabular} & Not examined in this study$^{**}$. \\ \hline
\end{tabular}
$^{*}$Laser prices are highly dependent on wavelength and exact characteristics. Given prices are just an indication for typical consumer price of a single device. \\
$^{**}$Refs. \cite{AfanadorDelgado2019,Zhang2022} show that an increase in dye concentration results in shorter nucleation times and smaller bubbles.
\end{table}

On the other hand, if the price and/or size of the device is of more importance, a CW laser would be preferred over a pulsed laser. Due to their lower power, they are smaller and more affordable. Although, portable pulsed lasers with moderate pulse energies, as the ones used in our studies, have become more widely available over the past decade. 

Furthermore, the use of optical fiber allows for a small handheld device, even in the case of a large laser source. The laser source, including electronics, can then be placed in a bench-top part. 
The choice of optical fiber has a large influence on bubble dynamics. For the pulsed laser, a smaller fiber reduces the required pulse energy. However, the smallest fiber (50~µm diameter) is limited in the range of bubble growth rate due to laser-induced damage of the fiber tip. For the CW laser, an increase in fiber size results in faster-growing bubbles, although it requires a more powerful laser (P~$>$~2W) to create a larger range of growth rates. 

Changing the dye concentration has a large influence on the bubble dynamics. For the pulsed laser, increase of dye concentration and thus absorption coefficient results reduces the required pulse energies, as discussion in Section~\ref{sec: pulsed laser}. This allows for the use of smaller and more affordable pulsed lasers. Simultaneously, a single device could create a large range of bubbles by changing the absorption coefficient of the liquid. The influence of the absorption coefficient on the bubbles generated by the CW laser have not been investigated in this study. However, previous studies showed an influence on the nucleation time and bubble size~\cite{AfanadorDelgado2019,Zhang2022}. Moreover, adding the dye into drug formulation might bear further toxicity and regulatory issues.  Future studies should focus on the exact influence of absorption coefficient on bubble dynamics.

\section{Conclusion}\label{sec: conclusion}

We compared the dynamics of bubbles generated  by two lasers with different timescales (ns and ms) inside an open-ended capillary. Our comparative study is the first to show the resulting bubble dynamics in the same fluidic confinement with two different laser types. We have shown that these lasers create bubbles of comparable growth rates proportional to the delivered energy. This linear increase is in agreement with previously found linear increase between the energy and the jet velocity. 

For the pulsed laser set-up, we found that the energy threshold for nucleation is proportional to the irradiated volume. This volume depends on the fiber core radius, beam divergence and absorption coefficient. Therefore, a decrease in irradiated volume, either by changing the fiber or increasing the absorption coefficient, resulting in a reduction of the energy threshold. This allows for the use of more affordable and/or less powerful pulsed laser sources while creating the same bubble. 

For the continuous-wave laser, we found an efficiency increase and better reproducibility compared to previous experiments with free-space optics. This is explained by a reduction in the number of interfaces and easier alignment. Furthermore, we show that the delivered energy can be controlled by the laser power and the fiber size. For each fiber, a decrease in laser power results in an increase in delivered energy, allowing the creation of faster-growing bubbles. Furthermore, a larger fiber results in an increase in delivered energy for fixed laser power due to a longer nucleation time. Therefore, the 200~µm fiber results in much larger and faster-growing bubbles than the 50 or 105~µm fiber, when the same laser power is used. 

A comparison between the two laser sources shows that the pulsed laser requires slightly less optical energy to create the same bubble growth rate, which we attribute to heat dissipation and a reduction in absorption coefficient during the CW laser heating. However, for the same average growth rate, bubbles generated by the CW laser grow for a longer time and are larger. 

Finally, we made a comparison between both methods in terms of practical usage. Since the delivered energy by the CW laser cannot be controlled directly, there is a larger deviation in the delivered energy compared to the pulsed lasers (8\% and 2\% respectively), which also results in a larger deviation in bubble growth rates. This would be unfavourable for laser-based jet injection, as it decreases the control over jet velocity and, therefore, injection depth. However, if this variation is within the allowable error industry standards, then the CW laser could be advantageous due to its smaller size and lower price compared to pulsed lasers.

\section*{Acknowledgements}
J.J.S. would like to thank dr. Dani\"el J\'auregui-V\'azquez and dr. Jose Alvarez Chavez for their help with the optical set-up of the CW laser. J.J.S and D.F.R. acknowledge the funding from the European Research Council (ERC) under the European Union’s Horizon 2020 Research and Innovation Programme (Grant Agreement No. 851630), and NWO Take-off phase 1 program funded by the Ministry of Education, Culture and Science of the Government of the Netherlands (No. 18844). J.K. and Ch.M. acknowledge the Innosuisse BRIDGE Proof of Concept grant funding. The authors are thankful for the insightful discussions with Dr. M. A. Quetzeri Santiago, D.L. van der Ven, K. Mohan and D. de Boer.

\section*{Competing interest}
The authors declare that they have no known competing financial
interests or personal relationships that could have appeared to influence the work reported in this paper.

\section*{CRediT authorship contribution statement}
\textbf{Jelle Schoppink:} Conceptualization, Methodology, Formal analysis, Investigation, Data Curation, Writing - Original Draft, Visualization
\textbf{Jan Krizek:} Conceptualization, Methodology, Writing - Review \& Editing
\textbf{Christophe Moser:} Conceptualization, Funding acquisition, Writing - Review \& Editing
\textbf{David Fernandez Rivas:} Conceptualization, Supervision, Project administration, Funding acquisition, Writing - Review \& Editing.

\printbibliography 

\begin{figure}[b!]
    \centering
    \includegraphics[width=0.7\textwidth]{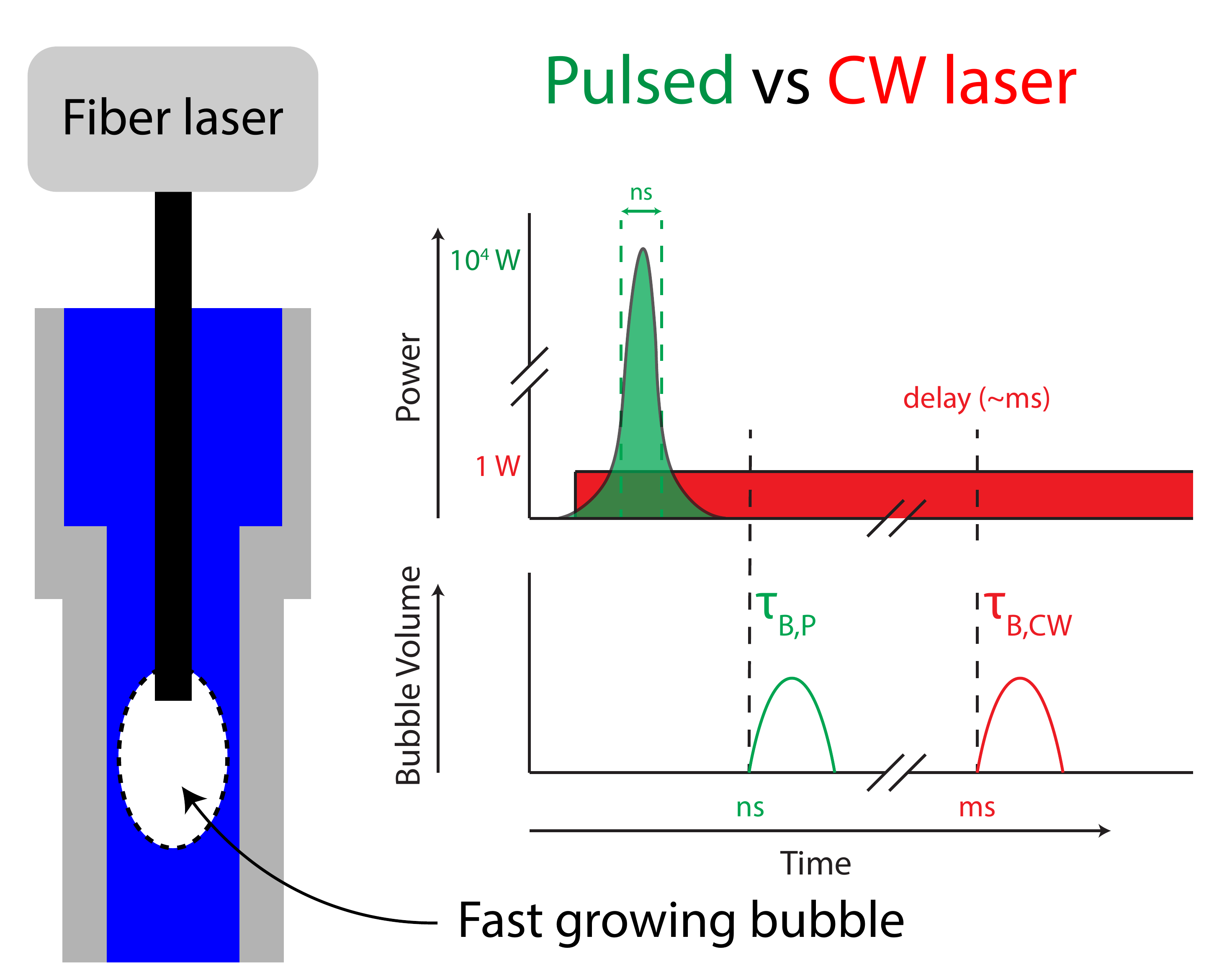}
    \captionsetup{width=1\linewidth}
    \caption{For Table of Contents Only}
    \label{fig: Graphical abstract}
\end{figure} 

\end{document}


\maketitle 

\section*{Appendix A: Allura Red AC absorption coefficient}\label{sec: Appendix}
\begin{figure}[h!]
\centering
  \includegraphics[width=.7\linewidth]{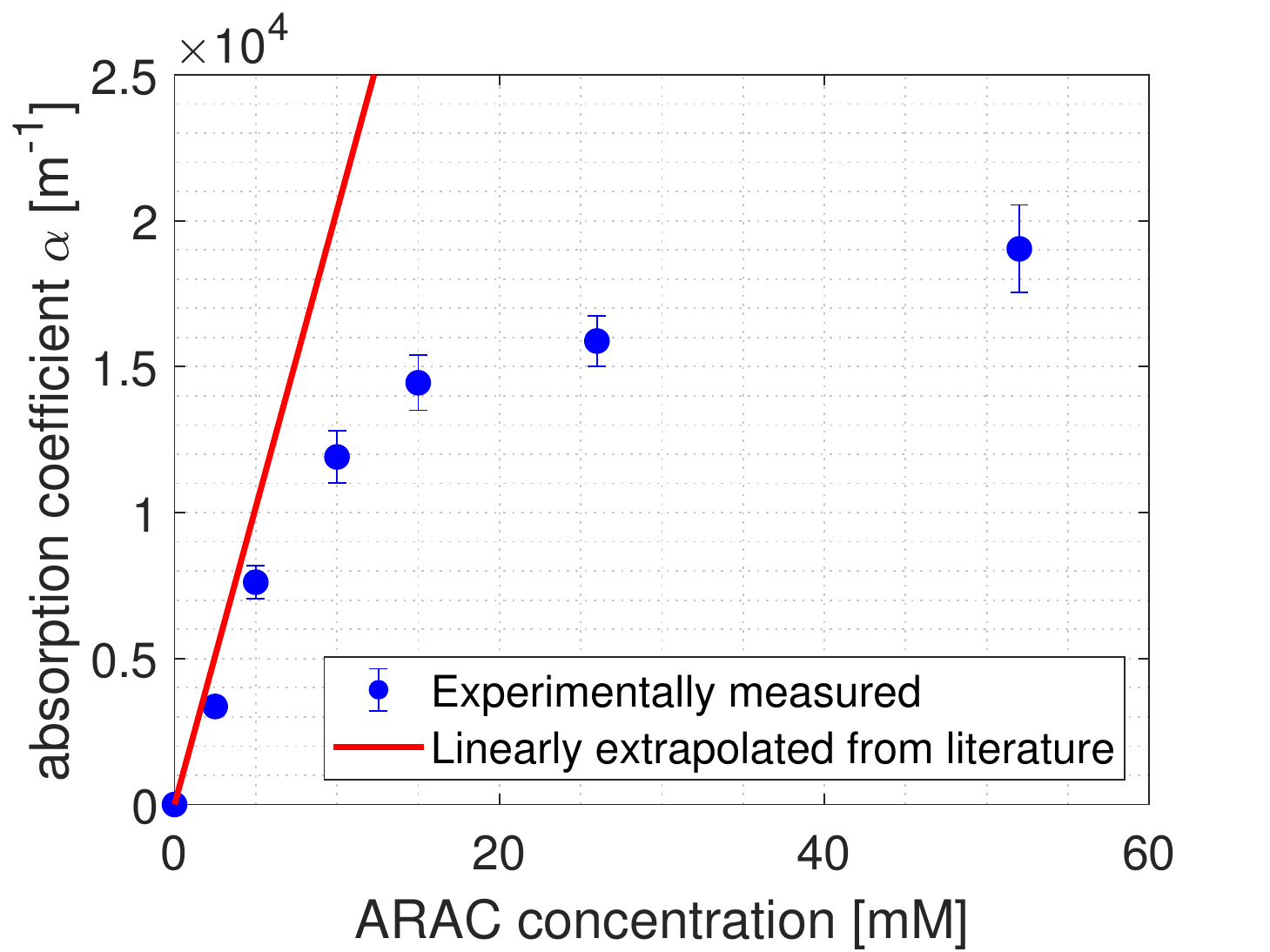}
  \caption{Measured absorption coefficients (blue symbols) for Allura Red AC (ARAC). The values are compared with an extrapolated value from literature (red curve)~\cite{Bevziuk2017,Garcia2017}. It can clearly be observed that the absorption coefficient does not behave linearly at these concentrations.}
  \label{fig: alpha_vs_concentration}
\end{figure}

The measured absorption coefficients for the Allura Red AC dye (ARAC, 80\%, Sigma Aldrich) are shown in Figure~\ref{fig: alpha_vs_concentration}. The values are calculated from the transmission of the 532~nm laser through a rectangular capillary with a 100~µm path length. The values are compared with reported values in literature~\cite{Bevziuk2017,Garcia2017}, which closely matches our measured values for  concentrations up to 5~mM. For larger concentrations, we note that the absorption coefficient does no longer increase linearly with the dye concentration.

\newpage
\section*{Appendix B: Calculation of irradiated volume}\label{sec: Appendix B}
\begin{figure}[h!]
    \centering
    \includegraphics[width = \linewidth]{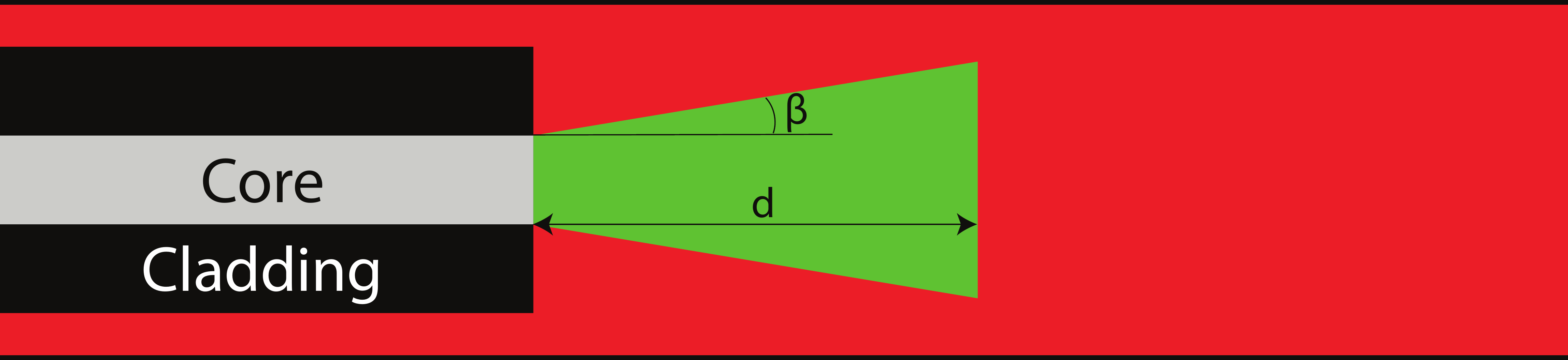}
    \caption{Schematic indicating the irradiated volume (green) of the liquid (red) near the fiber tip. Beam radius increases along the propagation axis due to the divergence. At the fiber tip, the radius is equal to the core radius. The irradiated volume is calculated by integrating the beam area from the fiber tip to the absorption length \textnormal{d}.}
    \label{fig: BeamRadDiv}
\end{figure}

As the fiber core radii R$_\textnormal{f}$ ($25\textnormal{~µm}<\textnormal{R}_\textnormal{f}<100\textnormal{~µm}$) are of similar size of the absorption length d ($52\textnormal{~µm}<\textnormal{d}<131\textnormal{~µm}$), the beam divergence $\beta$ cannot be neglected when calculating the irradiated volume. The beam divergence is calculated by the NA of the fiber and the refractive index of the liquid $n$ (see Figure \ref{fig: BeamRadDiv})
\begin{equation*}
    \beta = \arcsin{\frac{\textnormal{NA}}{n}} = \arcsin{\frac{\textnormal{0.22}}{1.33}} \approx 9.5\degree.
\end{equation*}
Then, the beam radius R along the propagation axis $x$ is calculated as
\begin{equation*}
    \textnormal{R}(x) = \textnormal{R}_\textnormal{f}+x\tan{\beta}.
\end{equation*}

The irradiated volume $V$ can be calculated by integrating the beam area ($\pi$R$(x)^2$) from the fiber tip ($x=0$) to the absorption length ($x=\textnormal{d}$)

\begin{align*}
    V = \pi\int_{0}^{\textnormal{d}}\textnormal{R}^2 dx = \pi\int_{0}^{\textnormal{d}}(\textnormal{R}_\textnormal{f}+x\tan{\beta})^2 dx = \pi\int_{0}^{\textnormal{d}}(\textnormal{R}_\textnormal{f}^2+2\textnormal{R}_\textnormal{f}x\tan{\beta}+x^2\tan^2{\beta})dx \\
    = \pi |\textnormal{R}_\textnormal{f}^2x + \textnormal{R}_\textnormal{f}x^2\tan{\beta}+\frac{1}{3}x^3\tan^2{\beta}|_0^\textnormal{d} = \pi (\textnormal{R}_\textnormal{f}^2\textnormal{d} + \textnormal{R}_\textnormal{f}\textnormal{d}^2\tan{\beta}+\frac{1}{3}\textnormal{d}^3\tan^2{\beta)}
\end{align*}
\newpage
\section*{Appendix C: CW bubble growth}\label{sec: Appendix C}
\begin{figure}[ht]
    \centering
\begin{minipage}{.49\textwidth}
  \centering
  \includegraphics[width=.9\linewidth]{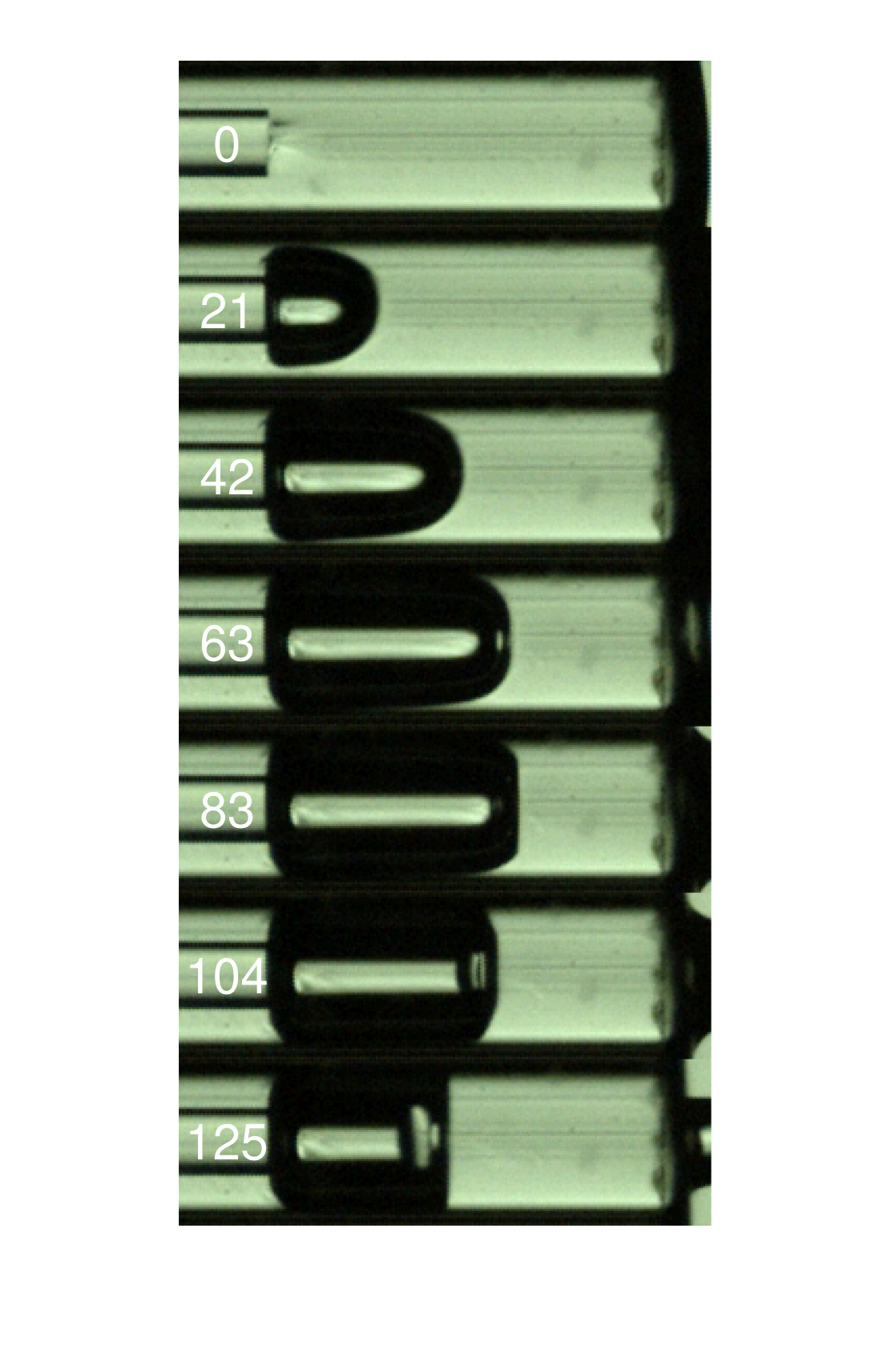}

\end{minipage} 
\begin{minipage}{.49\textwidth}
  \centering
  \includegraphics[width=.9\linewidth]{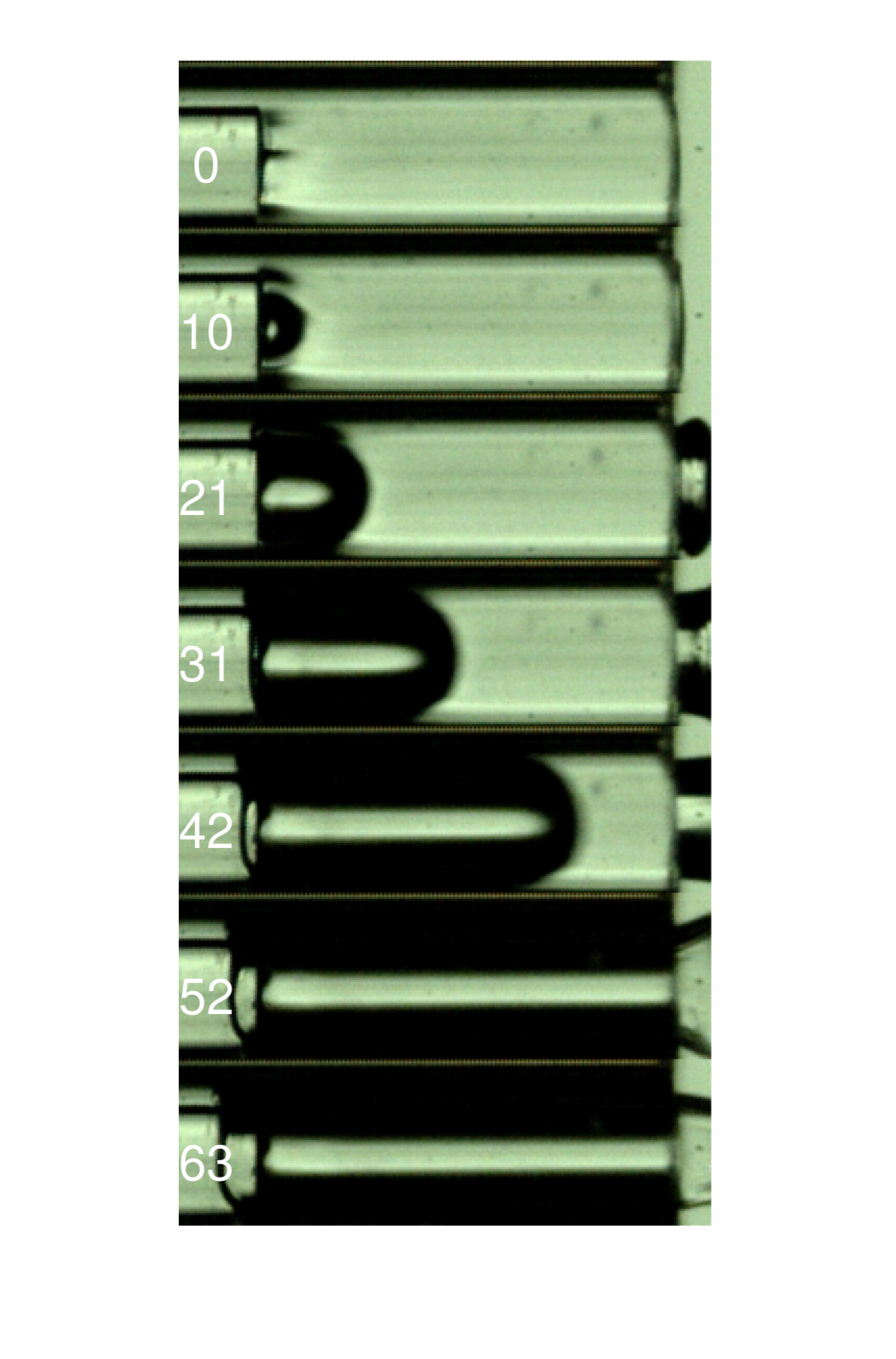}

\end{minipage}
    \caption{Snapshots of the bubble generated at a 105~µm (left)  and 200~µm fiber (right) by the CW laser with a power of approximately 0.7~W. The bubbles generated at the 200~µm fiber have a much larger energy and grow beyond the capillary edge and expel all the liquid. Therefore, there is no maximum bubble size or average growth rate observed.}
    \label{Appendix fig: CW 105 and 200}
\end{figure}

Figure~\ref{Appendix fig: CW 105 and 200} shows a few snapshots of the bubbles generated by the CW laser at the 105 and 200~µm fiber. For the 200~µm fiber (right snapshots), the bubble has a much larger energy, due to the longer nucleation time. Therefore, the bubble grows more explosively and always grows beyond the capillary edge. Therefore, no maximum bubble size or average growth rate could be obtained for the 200~µm fiber.

\printbibliography